\documentclass[showpacs,showkeys,superscriptaddress,groupedaddress,twocolumn,floatfix,prb]{revtex4-1}
\usepackage{graphicx}
\usepackage{amsmath}
\usepackage{amssymb}
\usepackage{amsbsy}
\usepackage{amscd}
\usepackage{color}
\usepackage{bbm}
\usepackage{braket}
\usepackage{bm}
\usepackage{siunitx}
\usepackage[normalem]{ulem} 
\usepackage{url}
\usepackage{footmisc}
\usepackage{tabularx} 
\usepackage[verbose,hypertexnames=false,bookmarksopenlevel=1,filecolor=blue,
linkcolor=blue,citecolor=blue, urlcolor=blue,pdfstartview=FitH,bookmarksopen,bookmarksnumbered,
colorlinks,plainpages=false,linktocpage]{hyperref}
\DeclareSymbolFont{mathbold}{OML}{cmm}{b}{it}
\newcommand{\K}{k_x^2-k_y^2}
\begin{document}
\title{Conserved Spin Quantity in Strained Hole Systems with Rashba
  and Dresselhaus Spin-Orbit Coupling} \author{Paul Wenk}
\email{paul.wenk@ur.de}
\author{Michael Kammermeier}
\author{John Schliemann} \affiliation{Institut f\"ur Theoretische
  Physik, Universit\"at Regensburg, D-93040 Regensburg, Germany}
\date{\today }
\begin{abstract}
  We derive an effective Hamiltonian for a (001)-confined
  quasi-two-dimensional hole gas in a strained zincblende
  semiconductor heterostructure including both Rashba and Dresselhaus
  spin-orbit coupling. In the presence of uniaxial strain along the
  $\left\langle110\right\rangle$ axes, we find a conserved spin
  quantity in the vicinity of the Fermi contours in the lowest valence
  subband. In contrast to previous works, this quantity meets
  realistic requirements for the Luttinger parameters. For more
  restrictive conditions, we even find a conserved spin quantity for vanishing
  strain, restricted to the vicinity of the Fermi surface.
\end{abstract}
\maketitle
\section{Introduction}
One of the most critical challenges for spintronic devices, as the
often mentioned spin-field-effect-transistor due to Datta and
Das\cite{:/content/aip/journal/apl/56/7/10.1063/1.102730}, lies in the
control of the carrier spin lifetime. The latter is limited by
the spin relaxation and dephasing processes in semiconductors. The
predominant mechanism of the spin relaxation in such devices is of
Dyakonov-Perel type\cite{dyakonov72_3}. To extend the application of
spintronic devices to the non-ballistic/diffusive regime with
spin-independent scattering, it is of particular interest to find
conditions for the electrons/holes in the semiconductors which result
in symmetries that correspond, according to Noether's
theorem\cite{Noether1918}, to the conservation of spin. These
symmetries enable to detect long-lived or even persistent spin states,
i.e., states which do not relax in time. In structurally confined
two-dimensional electron gases (2DEG) such persistent solutions have
been predicted for a special interplay between the
linear-in-momentum Dresselhaus\cite{Dresselhaus1955b} and
Bychkov-Rashba\cite{rashba_2,JPSJ.37.1325} spin-orbit coupling
(SOC) by J. Schliemann \textit{et al.},
Ref.~\onlinecite{Schliemann2003}, and were extended by
B.A. Bernevig \textit{et al.},
Ref.~\onlinecite{Bernevig2006}. These special states have been
later confirmed by means of optical
experiments\cite{Koralek2009,10.1038/nphys2383}. The first type of SOC
appears in semiconductors with broken inversion symmetry in the
crystal structure (bulk inversion asymmetry (BIA)), the second one
occurs when a structure inversion asymmetry (SIA) in consequence of an
asymmetric confining potential in the semiconductor heterostructure is
present.

In contrast to electron systems, the SOC is distinct and much more
complex for holes although the underlying fundamental mechanism,
described by the Dirac equation, is the same. Since the conduction
band, for most semiconductors, is an s-type energy band and the
valence band is of p-type, the qualitative variation comes from the
different total angular momentum, which is $j=3/2$ in the valence
band, giving rise to heavy and light holes (HH, LH) and the split-off
holes.  As the mixing of HH and LH strongly influences the SOC, the
reduction of dimensionality, like the 2D hole gases (2DHG) in
semiconductor heterostructures, has an immediate
effect.\cite{Rashba1988175} This is due to the fact that the size
quantization causes an energy separation between HH and LH states even
at a vanishing in-plane Bloch wave vector which affects the strength
of the HH-LH mixing and thus the SOC at finite $k$ vectors. As a
consequence, the magnitude of the SOC, especially the prefactor for
Rashba SOC, depends sensitively on the confinement, as will be
discussed in this paper. Conversely, the Rashba SOC in the lowest
conduction band is hardly affected by the size quantization. This is
due to the s-type character of the energy band and also since the SOC
is mainly determined by the energy gaps between the bulk bands. These
energy gaps, however, do not differ significantly if adding a
confinement. For a proper description of the hole system, compared to
electron systems, a substantially higher number of SOC terms is needed
and requires approximations for an analytical investigation at an
early stage.
 
Also, internal or external strain can yield significant consequences
to the hole band structure. The reason is that it introduces
additional couplings between the HH and LH bands, whereas - assuming a
semiconductor with a direct bandgap - the conduction band is only
indirectly affected due to the interaction with the strain-altered
valence band.\cite{YSunThompsonNishidaBook,
  :/content/aip/journal/jap/101/10/10.1063/1.2730561,
  Seiler77,PhysRevB.38.1806} Moreover, the cubic crystal structure of
the semiconductor has an imprint on the symmetry of the hole spectrum
as can be seen in the warping of the Fermi contours and these always
follow the strained crystal symmetry.

Nonetheless, despite their complexity hole systems offer opportunities
not available in electron systems and are particularly interesting for
practical device applications for several reasons.  First, the large
effective mass $m^*$ of holes compared to conduction band electrons
diminishes the kinetic term such that contributions from SOC become
more important. Second, the p-wave character of the HH and LH states
reduces the hyperfine interaction of the carrier spin with the nuclei.
This allows in principle for long spin relaxation/dephasing
times.\cite{1367-2630-12-4-043003, PhysRevB.80.035325} Another
important aspect is the strength of the SOC in hole systems which can
reach several meV in the splitting as, e.g., shown in GaAs/AlGaAs
heterostructures\cite{PhysRevLett.107.216805, PhysRevB.77.125312}.
All these features of p-type systems facilitate a very effective
manipulation of carrier spins and, hence, motivate further studies of
hole gases in semiconductors as the one presented here.  Moreover,
since in a 2DEG the conserved spin quantities are always limited by a
$k$-cubic Dresselhaus contribution the question arises whether this is
also the case for the 2D hole systems.

An appealing continuation of the findings on spin-preserving
symmetries in electron systems is the analysis of persistent spin
states in hole systems as done recently in
Refs.~\onlinecite{PhysRevB.89.161307,PhysRevB.90.115306}. However,
these publications presuppose materials with strongly restricted and
unusual band structures. Following
Ref.~\onlinecite{PhysRevB.90.115306}, a strainless sample with both
Rashba and Dresselhaus SOC allows for the existence of a persistent
spin helix (PSH) in a 2DHG only in the case of a vanishing Luttinger
parameter $\gamma_3$ with $\gamma_1>0$ and
$\gamma_2>0$.\cite{PhysRevB.90.115306} Most of the semiconductors can
only be properly described using a band model where
$\gamma_2<\gamma_3$\cite{:/content/aip/journal/jap/89/11/10.1063/1.1368156,
  YuCardonaBook, winklerbook}(App.~\ref{luttinger_param_relation}),
though.  In Ref.~\onlinecite{PhysRevB.89.161307} the PSH was found in
the presence of finite strain and Rashba SOC where the condition for
the Luttinger parameters is restricted by a different, however, also
unusual condition $\gamma_2=-\gamma_3$\footnote{A negative value of
  $\gamma_2/\gamma_3$ appears, e.g., when the lowest conduction band
  is not an s-type band as it is the case in
  diamond\cite{PhysRevB.50.18054}. However, for diamond one finds only
  $\gamma_2/\gamma_3\approx -0.16$\cite{PhysRevB.50.18054}.}.

Another approximation which is often applied is to drop all invariants
in the bulk Hamiltonian which lead to BIA and have relatively small
expansion coefficients. This procedure is justified in bulk
systems. In 2DHGs, this approach leads to a model Hamiltonian with
both Rashba and Dresselhaus SOC being essentially cubic in
momentum\cite{PhysRevLett.95.076805, PhysRevB.90.115306}, in contrast
to 2DEGs where the dominant Rashba term is linear and the Dresselhaus
term linear and cubic in momentum.\cite{winklerbook} However, recent
publications\cite{PhysRevLett.104.066405, PhysRevB.89.075430} which
are related to the seminal paper by Rashba and Sherman,
Ref.~\onlinecite{Rashba1988175}, show that the relevance of the linear
Dresselhaus terms in 2DHG has been underestimated: the abovementioned
estimations are thus questionable and fail at least for the standard
compound GaAs.

In this paper, we present conditions for a wider range of
semiconductors including strain, linear and cubic Dresselhaus and
Rashba SOC under which conserved spin quantities can be found.
\\
This paper is organized as follows. In the next section, we derive an
effective two-dimensional heavy/light hole like Hamiltonian including
strain, Dresselhaus and Rashba SOC.  In Sec.~\ref{sec:conserved}, we
derive the conditions for the existence of a conserved spin quantity.
Thereby we discuss its realizability and apply our findings to a
prominent compound, InSb. Finally, we summarize our results.
\section{The Model}
The aim of our investigation is to identify the appropriate interplay
between BIA (Dresselhaus SOC), a confining potential
$V({\bf r})$ (build-in and/or external) causing SIA, and strain
(either externally imposed using, e.g., the piezoelectric effect or
induced by the epitaxial growth process) which gives rise to a
conserved spin quantity in the hole system. To find analytic
conditions we derive an effective HH/LH like $2 \times 2$ model,
depending on the character of the up-most valence band.  Starting
point is a $4 \times 4$ model which is derived form the extended
Kane model and includes the Luttinger
Hamiltonian\cite{Luttinger1956}.
\subsection{Effective $4 \times 4$ Hole Hamiltonian}\label{sec:4x4}
Hereafter, we choose the coordinates to be
$\hat{x}\parallel [100]$, $\hat{y}\parallel [010]$ and
$\hat{z}\parallel [001]$.
We use the Luttinger parameters $\gamma_i$, the bare electron mass
$m_0$ and elementary charge $e>0$, the electric field
$\mathcal{E}_z$, the total angular momentum $\mathbf{J}$ for $j=3/2$
and the symmetrized anticommutator $\left\{A,B\right\}=\left( AB+BA
\right)/2$.

The applied model which we use as a starting point for the
investigation is an effective $4\times 4$ Hamiltonian given by
\begin{align}
  \mathcal{H}={}&\mathcal{H}_\text{L}+\mathcal{H}_\text{BIA}
  +\mathcal{H}_\text{S}+V.
\label{bulk}
\end{align}
The first term represents the Luttinger
Hamiltonian for III-V semiconductors
\begin{align}
  \mathcal{H}_{\text{L}} ={}& -\frac{\hbar^2}{2m_0}\Bigg( \gamma_1
  \mathbf{k}^2 -2 \gamma_2 \left[ \left(J_x^2-\frac{1}{3}\mathbf{J}^2
    \right) k_x^2
    +\text{c.p.} \right]\notag\\
  &- 4 \gamma_3\left[ \{J_x,J_y\}\{k_x,k_y\}+\text{c.p.}\right]
  \Bigg),\label{luttinger}
\end{align}
where \textit{c.p.} denotes the \textit{cyclic permutation} of the
preceding indices.
The second term $\mathcal{H}_\text{BIA}$ accounts for the Dresselhaus
SOC and is decomposed by the theory of invariants as\cite{winklerbook}
\begin{align}
  \mathcal{H}_\text{BIA}={}&\frac{2}{\sqrt{3}}\,C_k\left(k_x\{J_x,J_y^2-J_z^2\}
    +\text{c.p.}\right)\notag\\
  {}  &+b_{41}^{8v8v}\left(\{k_x,k_y^2-k_z^2\}J_x+\text{c.p.}\right)\notag\\
  {}&+b_{42}^{8v8v}(\{
  k_x,k_y^2-k_z^2 \}J_x^3+\text{c.p.})\notag\\
  {}&+b_{51}^{8v8v}( \{ k_x,k_y^2+k_z^2 \}\{ J_x,J_y^2-J_z^2
  \}+\text{c.p.} )\notag\\
  {}& +b_{52}^{8v8v}( k_x^3\{ J_x, J_y^2-J_z^2 \} + \text{c.p.}).
\label{bia}
\end{align}
It includes a $k$-linear term proportional $C_k$ which is usually
rather small\cite{PhysRevB.38.1806,winklerbook} and thus often not
considered\cite{PhysRevB.89.075430, PhysRevLett.95.076805,
  PhysRevB.90.115306, Rashba1988175}.  An exemplary comparison of the
remaining cubic contributions with the extended Kane model in
App.~\ref{dominantInvariants} shows that in a bulk system the term proportional
to $b_{41}^{8v8v}$ is the most relevant one.  However, in 2D size
quantization leads to additional linear terms that may become
significant in certain parameter
regimes\cite{PhysRevB.89.075430,Rashba1988175,PhysRevLett.104.066405}. We
note that there are discrepancies in the perturbative determination of
the coefficients $b_i^{8v8v}$ as outlined in App.~\ref{dominantInvariants}.

Furthermore, the effect of strain is described by the Bir-Pikus strain Hamiltonian
$\mathcal{H}_\text{S}$\cite{BirPikusSymmetryStrain1974,PhysRevB.89.161307}:
\begin{align}
  \mathcal{H}_\text{S}={}&\sum_{i}\left(a \, \epsilon_{ii}+b \;
    \epsilon_{ii} \, J_{i}^2+ d \, \sum_{j,\; j \neq i}\epsilon_{ij}
    \, \{J_{i},J_{j}\}\right).\label{strain}
\end{align}
We assume the strain, with $\epsilon_{ij}$ being the
symmetric strain tensor, to be uniaxial in-plane or biaxial
in-plane. As a consequence, an in-plane strain with the tensor
components $\epsilon_{xx}$, $\epsilon_{yy}$, $\epsilon_{xy}$ yields
only one extra strain component, $\epsilon_{zz}$. Thus, we set
$\epsilon_{xz}=\epsilon_{yz}=0$ in Eq.~(\ref{strain}). One should
stress that slightly different notations occur in the literature
for the deformation potentials $a$, $b$ and $d$. Some relations
between different definitions are listed in
App.~\ref{sec:deformation_potentials}. Here, we follow
Ref.~\onlinecite{PhysRevB.89.161307}. Thus, the potentials $b$ and $d$
are positive whereas $a$ can be positive or negative.

Eventually, the potential $V=V_{\text{E}}+V_\text{c}$ includes the confining
potential $V_\text{c}(z)$ and the external potential
$V_{\text{E}}(z)$. 
The latter causes structure inversion asymmetry (SIA) which induces Rashba SOC. 
We assume the potential $V$ to depend only
on the $z$-coordinate, which is pointing in the growth direction $[001]$ of the
semiconductor heterostructure. More explicitly,
\begin{align}
  V_\text{E}(z)=&\mathbbm{1}_{4\times4}\cdot e\, \mathcal{E}_z
  z,\label{asym}
\end{align}
and an infinite square well of width $L$ with
\begin{align}
V_{\text{c}}(z)= \mathbbm{1}_{4\times4}\cdot
\begin{cases}
0 & \text{for}\; z \in \left[0,L\right],\\
\infty & \text{otherwise}.
\end{cases}
\label{conf}
\end{align}

Notice that the discontinuity of the potential $V_{\text{c}}$ can
result in non-hermitian matrix elements of $k_z^3$ if not taken
care. This problem can be resolved by a regularization procedure as
shown in Ref.~\onlinecite{PhysRevB.90.195410}.  Furthermore,
contributions due to boundary effects which result from the presence
of heterointerfaces are assumed to be small. Such an interface can
allow for additional HH-LH mixing.\cite{IvchenkoSuperlattices,
  PhysRevB.54.5852, PhysRevB.62.10364, PhysRevLett.104.066405} A
possible alternation of spin relaxation due to interface effects will
be discussed elsewhere.

\subsection{Effective $2 \times 2$ Model for the First Subband}
In the following, we choose the basis states in such a way that the
upper left block represents the HH the lower right block the LH
subspace:
\begin{align}
  \mathcal{H}=
  \begin{pmatrix}
    \mathcal{H}_\text{HH}  & \mathcal{H}_\text{HH-LH}\\
    \mathcal{H}_\text{LH-HH}&\mathcal{H}_\text{LH}
  \end{pmatrix}.\label{blocks}
\end{align}
The confinement in $z$-direction allows for a further simplification
of the model to an effective $2\times 2$ Hamiltonian using
quasi-degenerate perturbation theory (L\"owdin's partitioning). The
full Hamiltonian $\mathcal{H}$ is separated into two parts
\begin{align}
\mathcal{H}=\mathcal{H}_0+\mathcal{H}^\prime,\label{H_prime}
\end{align}
according to App.~\ref{secloewdin}. The partition is, in
general, not uniquely defined and different ways of splitting
$\mathcal{H}$ are possible. For the given system, a meaningful
decomposition is the one which allows a projection on the subspace of
a particular HH or LH like subband. Here, we select $\mathcal{H}_0$ to
contain the diagonal elements of
$\mathcal{H}_\text{L}+\mathcal{H}_\text{S}+V_\text{c}(z)$ at
$k_x=k_y=0$, and $\mathcal{H}^\prime$ is treated as a perturbation
with respect to the appropriate inverse splitting
$1/\Delta_\text{hl}$, ($1/\Delta_\text{hh}$)
between a HH like and a LH(HH) like subband. The energy splitting
$\Delta_\text{hl}$ is due to both the spatial confinement in the [001]
direction and the imposed strain.

According to the confinement, the eigenstates of $\mathcal{H}_0$ are
given by $\ket{j,m_{j}}\ket{n}$, the product of the eigenstates of the
total angular momentum $\mathbf{J}$ with $j=3/2,m_j=\pm1/2$ for LH and
$m_j=\pm3/2$ for HH and the subband index of $z$-quantization $n \in
\mathbbm{N}^*$. The eigenfunctions of the quantum well
in position space are given by $\braket{z|n}=\sqrt{2/L} \sin(z \,n
\,\pi/L )$ which lead to the matrix elements of the $k_i$ and $z$
operators given by
\begin{align}
\braket{n|k_z|l} ={}& \frac{2 i n l \left((-1)^{l+n}-1\right)}{L\left(n^2-l^2\right)}(1-\delta_{nl})\label{kz},\\
\braket{n|k_z^2|l} ={}& \left(\frac{\pi n}{L}\right)^2 \delta_{nl}, \label{kz2}\\
\braket{n|z|l}={}&
\begin{cases}
\frac{L}{2} & \text{for } n=l,\\
\frac{4 n l L\left((-1)^{l+n}-1\right)}{\pi^2\left(l^2-n^2\right)^2} &
\text{otherwise.}
\end{cases}
\label{z}
\end{align}
The eigenenergies of $\mathcal{H}_0$ (twofold degenerate) according to
the sub-spaces are given by
\begin{align}
E_\text{HH}(n) ={}& -\frac{\hbar^2}{2 m_0}(\gamma_1-2\gamma_2)\braket{k_z^2}n^2\label{EHH_n},\\
E_\text{LH}(n) ={}& -\frac{\hbar^2}{2 m_0}(\gamma_1+2\gamma_2)\braket{k_z^2}n^2+\delta,\label{ELH_n}
\end{align}
where
$\delta=b\left(\epsilon_{xx}+\epsilon_{yy}-2\epsilon_{zz}\right)$
corresponds to the diagonal shear strain tensor and
$\braket{k_z^2}\equiv\braket{1|k_z^2|1}$. For simplification we
subtracted an overall constant energy shift of $\Delta E=-\sum_{i}(a+3
b/4) \epsilon_{ii}-(3/2)b\epsilon_{zz}$.  It becomes apparent that
strain effects as well as the confining potential lift the degeneracy
of HH and LH bands at $k_\parallel\equiv \sqrt{k_x^2+k_y^2}=0$.

In the following, the in-plane strain contribution will be rewritten,
according to Ref.~\onlinecite{PhysRevB.89.161307}, defining an
in-plane strain amplitude $\beta$ and orientation $\theta$
as\footnote{Note that the definition in Eq.~(\ref{beta_def}) differs
  from the one given in Ref.~\onlinecite{PhysRevB.89.161307} in the
  term proportional to $d$ by a factor of $1/\sqrt{3}$.}
\begin{align}
  \beta^2 e^{2i\theta}:={}& \frac{m_0}{\hbar^2\gamma_3}(b\left( \epsilon_{xx}-\epsilon_{yy}
    \right)+2 i d \epsilon_{xy}).\label{beta_def}
\end{align}
The effective Hamiltonian for the lowest subband is now derived by
quasi-degenerate perturbation theory (see
App.~\ref{secloewdin}). Thereby, the lowest subband can be either a
HH like or a LH like subband, depending on the magnitude and type of
strain. Up to third order in the energy splitting it yields
\begin{align}
  \mathcal{H}_{\text{eff}} \approx{}& \sum_{p=0}^{3} \mathcal{H}^{(p)}\\
  ={}&\sum_{p=0}^{3} \left(E_\text{kin}^{(p)}+V_\text{eff}^{(p)}\right)
  \cdot
  \mathbbm{1}_{2\times2}+\boldsymbol{\Omega}^{(p)}\cdot\boldsymbol{\sigma}
\label{perturb}
\end{align}
with the Pauli matrices $\sigma_i$ and the $\mathcal{H}^{(p)}$, where
the superscript $(p)$ indicates the order in the perturbation
(App.~\ref{secloewdin}). Additionally, we will neglect terms of the
order of $\mathcal{O}(k^4)$.

We would like to mention some properties of the perturbation theory
first. Concerning the decomposition of the Hamiltonian, it should be
stressed that if we instead choose
$\mathcal{H}_0=\mathcal{H}_\text{L}+\mathcal{H}_\text{S}$, that is,
without the confinement potential, we can make use of the
non-commutativity of the momentum operator $k_z$ and the position
operator $z$ to derive a finite Rashba coefficient as done in
Refs.~\onlinecite{winklerbook,PhysRevB.89.161307}. However in this
case, the energy splitting $\Delta_\text{hl}$ cannot be a result of
the subband quantization, but only have strain as an origin. This
approximation may be justified if the strain splitting is much larger
than the subband splitting. Yet, since we consider a quasi 2D system,
the subband splitting is an essential effect and we are thus to choose
the partitioning as described above.

An important observation of the perturbation presented in this paper
is that a finite Rashba spin-orbit (SO) field, resulting from the coupling between
different subbands, can only be obtained by third or higher order
perturbation theory. This is due to the fact that  the
diagonal elements of $\mathcal{H}^\prime$, Eq.~(\ref{H_prime}), and
thus $\mathcal{E}_z$, are not yet involved in the second order, according
to Eq.~(\ref{perturbationformula3}). In addition, it will be shown in
the following that given a HH(LH) like ground state it is necessary to
include, in addition to the first LH(HH) like subband, also the
second HH(LH) like subband in the perturbation procedure to yield a
finite contribution due to Rashba SOC.

Concerning the significance of the various Dresselhaus contributions
in Eq.~(\ref{bia}) we mentioned earlier, Sec.~\ref{sec:4x4}, that in
the bulk system keeping only the cubic term proportional
$b_{41}^{8v8v}$ is a good approximation.  This is due to its large
value compared to the other cubic terms.\cite{winklerbook} However,
the size quantization causes an additional linear Dresselhaus
contribution for HHs that would not appear if only the term
proportional $b_{41}^{8v8v}$ was considered.  For the light holes, the
situation is different as the term proportional $b_{41}^{8v8v}$
already yields a linear term which clearly dominates over the
remaining linear contributions.  As a result, we will take into
account the effect of the $k$-linear terms generated in the first
order perturbation by the terms proportional $C_k$, $b_{42}^{8v8v}$
and $b_{51}^{8v8v}$ only in case of a HH like ground state and neglect
them in case of a LH like ground state.  The coefficient
$b_{52}^{8v8v}$ yields only a small cubic term which can be
disregarded.  Based on its large value, for higher order BIA
corrections we incorporate solely terms proportional to
$b_{41}^{8v8v}$.


Depending on the nature of the strain, the splitting can lead to
either a lowest HH like or LH like subband. If we, e.g., specify to the
case of uniaxial compressive stress in [110] direction, we obtain
$\delta<0$ since $\epsilon_{xx}=\epsilon_{yy}<0$ and
$\epsilon_{zz}>0.$\cite{:/content/aip/journal/jap/101/10/10.1063/1.2730561}
Therefore, the splitting between HH and LH is enhanced and the topmost
subband is HH for arbitrary material since $\gamma_i>0$. Consequently,
we have to distinguish two cases:\\

\subsubsection{System with a Heavy-Hole Like Ground State}
Assuming the ground state being HH like and applying third order
L\"owdin perturbation theory, we get for the part of the Hamiltonian
which is proportional to unity in spin space, according to
Eq.~(\ref{perturb}),
\begin{align}
  E_{\text{kin,HH}} ={}& E_{\text{kin,+}},\\
  V_{\text{eff,HH}} ={}& V_{\text{eff},+}(\Delta_{\text{h1,h2}}),
\end{align}
where we defined
\begin{widetext}
  \begin{align}
    E_{\text{kin},\pm} = -\frac{\hbar^2}{2
      m_0}\left(\gamma_1\pm\gamma_2\right)k_\parallel ^2
    \pm\frac{1}{\Delta_{\text{h1,l1}}}\left\{\frac{3}{4}
      \left(b_{41}^{8v8v}\braket{k_z^2}\right)^2 k_\parallel^2
      +\frac{3\hbar^4}{2m_0^2} \beta^2 \gamma_3^2
      \left[\frac{\gamma_2}{\gamma_3} \left(\K\right)\cos(2 \theta)+2
        k_x k_y \sin (2 \theta )\right]\right\}\label{Ekin_HH}
  \end{align}
  and
  \begin{align}
    V_{\text{eff},\pm}(\Delta)= \sum_{i}\left(a+ \frac{3}{4}b\right)
    \epsilon_{ii}+\frac{3}{2}b\epsilon_{zz}-\frac{\hbar^2
      \braket{k_z^2} }{2 m_0}\left(\gamma_1\mp 2\gamma_2\right)+
    \frac{e\mathcal{E}_zL}{2} \pm\frac{3\hbar^4}{4
      m_0^2\Delta_{\text{h1,l1}}} \beta^4 \gamma_3^2+\frac{256 L^2 e^2
      \mathcal{E}_z^2}{81\Delta}.
  \end{align}
The energy gaps are given by
$\Delta_\text{ln,hm}=E_{\text{LH}}(n)-E_{\text{HH}}(m)$ and
$\Delta_\text{hn,lm}$ analogously.
The effective vector field $\boldsymbol{\Omega}$ due to Rashba and
Dresselhaus SOC, which is modified by the presence of strain, yields
\begin{align}
  \boldsymbol{\Omega}_\text{HH}=\boldsymbol{\Omega}_+\label{sofHH}
\end{align}
with components given by
  \begin{subequations}
    \begin{align}
      \Omega_{x,\pm} = {} & \lambda_{\text{D},\pm}\Bigg{\{} k_x
      k_y^2\left(\gamma_2 \mp 2\gamma_3\right)-k_x^3\gamma_2
      +\beta^2\gamma_3\left[\pm k_y \left(
          \frac{k_x^2}{\braket{k_z^2}}-1 \right) \sin(2 \theta) +k_x
        \left( \frac{k_y^2}{\braket{k_z^2}}-1 \right) \cos (2
        \theta)\right]\Bigg{\}} \notag\\ &
      +\eta_{\pm}k_x+\lambda_{\text{R},\pm}\left[\pm \left(\gamma_2\pm
          2\gamma_3\right)k_x^2k_y\mp \gamma_2
        k_y^3+\beta^2\gamma_3\left(\pm
          k_y \cos(2\theta)+k_x\sin(2\theta)\right)\right], \\
      \Omega_{y,\pm}={}&\lambda_{\text{D},\pm}\Bigg{\{}\pm k_x^2
      k_y\left(\gamma_2\mp 2\gamma_3\right)\mp k_y^3\gamma_2
      +\beta^2\gamma_3\left[k_x \left( \frac{k_y^2}{\braket{k_z^2}}-1
        \right) \sin(2 \theta)
        \mp k_y \left( \frac{k_x^2}{\braket{k_z^2}}-1 \right) \cos (2 \theta)\right]\Bigg{\}}\notag\\
      {}&\pm\eta_{\pm}k_y\pm\lambda_{\text{R},\pm}\left[\pm\left(\gamma_2\pm
          2\gamma_3\right)k_xk_y^2\mp \gamma_2 k_x^3
        +\beta^2\gamma_3\left(k_y
          \sin(2\theta)\mp k_x\cos(2\theta)\right)\right],\\
      \Omega_{z,\pm}={}&0.
    \end{align}
  \end{subequations}
\end{widetext}
Here, the Dresselhaus coefficients $\eta_{\pm}$, $\lambda_{\text{D},\pm}$ and the Rashba
coefficient $\lambda_{\text{R},\pm}$ are given by
\allowdisplaybreaks
\begin{align}
\eta_{+} ={}&-\frac{\sqrt{3}}{2}C_k-\frac{3}{4}\left(b_\text{42}^{8v8v}+b_\text{51}^{8v8v}\right)\braket{k_z^2} \label{etaHH}\\
\eta_{-} ={}& -b_\text{41}^{8v8v}\braket{k_z^2}\label{etaLH}\\
  \lambda_{\text{D},\pm} ={}& \pm\frac{3\hbar^2}{2m_0\Delta_{\text{h1,l1}}}b_\text{41}^{8v8v}\braket{k_z^2},\label{lambdaD}\\
  \lambda_{\text{R},+} ={}& \frac{128 \hbar^4 e\mathcal{E}_z\gamma_3}{9 \pi^2 m_0^2 \Delta_\text{h1,l1}\Delta_\text{h1,h2}},\label{lambdaRP}\\
  \lambda_{\text{R},-} ={}& \frac{128 \hbar^4
    e\mathcal{E}_z\gamma_3}{9 \pi^2 m_0^2
    \Delta_\text{l1,h1}\Delta_\text{l1,l2}}.\label{lambdaRM}
\end{align}
The index $(\pm)$ distinguishes the case of a system with a HH like
ground state $(+)$ from the one with a LH like ground
state $(-)$.

Notice that given a vanishing $\gamma_3$ the dominant contribution due
to Rashba SOC vanishes, too. Only a contribution as a result of the
coupling to conduction bands is left, which is of higher than third
order: In the bulk system, for most semiconductors, the dominant
invariant in the extended Kane model\cite{Kane1957249}, which is
present due to SIA, is given by\cite{winklerbook}
\begin{align}
\mathcal{H}_{8v8v}^r=r_{41}^{8v8v}((k_y\mathcal{E}_z-k_z\mathcal{E}_y)J_x
+\text{c.p.}).\label{H8v8v}
\end{align}
If the bulk system is reduced to an effective two-dimensional system,
the according counterpart of this term in an effective $2\times 2$
Hamiltonian can be calculated using L\"owdin perturbation theory as done
above, keeping the factor $r_{41}^{8v8v}$ unchanged. However, as
mentioned in Ref.~\onlinecite{PhysRevB.90.115306}, for a HH like
ground state this resulting term is of higher order than the one given
proportional to $\lambda_\text{R}$ (although represented by the same
invariants). This can be understood by recalling the root of the
coefficient $r_{41}^{8v8v}$: It is the coupling between valence and
conduction bands. In contrast, if a confinement is present, the
contribution resulting from Rashba SOC in the effective HH system is
dominated by the splitting between HH and LH like subbands.  In the
case of a LH like ground state an additional $k$-linear Rashba term
proportional to $r_{41}^{8v8v}$ appears already in third order
perturbation theory. The angular momentum matrix $J_x$ is zero in
the HH subspace but has finite matrix elements in the LH
subspace. Since the prefactor $r_{41}^{8v8v}$ contains terms which are
inversely proportional to the band gap\cite{winklerbook}, this
contribution can be neglected since we assume the conduction bandgap
to be much larger than the subband splitting. The contribution
stemming from BIA has a different nature: The parameter
$b_{41}^{8v8v}$, which is connected with the invariant
$(\{k_x,k_y^2-k_z^2\}J_x+\text{c.p.})$ in the Kane model, is mainly
defined through the valence band $\Gamma_{8v}$ and conduction band
$\Gamma_{6c}$ gap $E_0$. Thus, it is hardly affected by the subband
quantization. Moreover, in contrast to the Rashba contribution, the
corresponding Dresselhaus term in the confined system appears already in
second order of the applied perturbation. Hence, we also neglect higher
order contributions due to BIA.
\subsubsection{System with a Light-Hole Like Ground State}
According to Eqs.~(\ref{EHH_n}) and (\ref{ELH_n}), on condition that
$\delta>2\hbar^2 \pi^2 \gamma_2/(m_0 L^2)$ the ground state of the
valence band is the first LH like subband. As in the case of a HH like
ground state, we do not obtain a $z$-component in the effective SO
field. However, in first order L\"owdin perturbation theory,
Eq.~(\ref{perturbationformula1}), we obtain an additional linear term
$\eta_-$ and a cubic term $\boldsymbol{\Omega}_b$ proportional to
$b_{41}^{8v8v}$.  Furthermore, terms appear in third order which
couple the electric field $\mathcal{E}_z$ with the Dresselhaus term
proportional to $b_{41}^{8v8v}$, App.~\ref{sec:lh-like-valence}.
Since the SOC is a small correction, these terms are much smaller than
the one not mixing both factors.  Thus, according to the previous case
of a HH like ground state, we have
\begin{align}
  E_{\text{kin,LH}} ={}& E_{\text{kin},-},\\
  V_{\text{eff,LH}} ={}& V_{\text{eff},-}(\Delta_{\text{l1,l2}})+\delta.
\end{align}
The effective vector field $\boldsymbol{\Omega}$ due to Rashba and
Dresselhaus SOC yields
\begin{align}
  \boldsymbol{\Omega}_\text{LH}=\boldsymbol{\Omega}_-+\boldsymbol{\Omega}_b\label{sofLH}
\end{align}
with the additional term 
\begin{align}
  \boldsymbol{\Omega}_b =b_{41}^{8v8v}\left\{k_x k_y^2,- k_x^2 k_y,0\right\}^\top
.\label{omega_b}
\end{align}

\subsection{Summarized Results}
In summary, by developing an effective $2 \times 2$ model for a 2DHG
we worked out the dominant contributions due to strain ($\beta$,
$\theta$), Rashba ($\lambda_R$) and Dresselhaus SOC ($\eta$,
$\lambda_D$) to be considered in Eq.~(\ref{perturb}).  The interplay
between strain and Rashba or Dresselhaus SOC yields additional terms
that are linear respectively linear as well as cubic in momentum.
Thereby, we find that in contrast to the Rashba contribution the
according Dresselhaus term in the confined system appears already in
second order L\"owdin perturbation theory.  In respect of finding a
conserved spin quantity we extracted the effective vector fields
$\boldsymbol{\Omega}_\text{HH}$ for a HH like ground state and
$\boldsymbol{\Omega}_\text{LH}$ for a LH like ground state in
Eqs.~(\ref{sofHH}) and (\ref{sofLH}).  The fields cover a wide
parameter space. In the next section, this will allow for identifying
conserved spin quantities which do not require
parameter-configurations which are difficult to realize in real
materials (e.g., $\gamma_3=0$ in Ref.~\onlinecite{PhysRevB.90.115306},
$\gamma_2=-\gamma_3$ in Ref.~\onlinecite{PhysRevB.89.161307}). Thus,
it facilitates the detection of long-lived spin states in experiments.
\section{Conserved Spin Quantity}\label{sec:conserved}
Following the analysis of Ref.~\onlinecite{Schliemann2003}, our goal
is to identify a conserved quantity $\Sigma$ which is
directly connected with $k$-independent eigenspinors. The general
ansatz is
\begin{equation}
  \Sigma=s_0 \mathbbm{1}_{2\times2}+\mathbf{s}\cdot \boldsymbol{\sigma}.
  \label{cq}
\end{equation}
For this quantity to be conserved, it has to fulfill the relation
$\left[\Sigma,\mathcal{H}_\text{eff}\right]=0$ which is
true for
\begin{equation}
  \Omega_x s_z=\Omega_y s_z=0 \quad \wedge \quad \Omega_y s_x-\Omega_x s_y=0.
  \label{condition}
\end{equation}
We are going to prove that one can find two solutions of this
problem given by
\begin{equation}
  \Sigma_\xi=\sum_{{\bf k},k=k_{\parallel F}}\sum_{\alpha\beta}c^\dagger_{{\bf
      k}\alpha}(\sigma_x+\xi\sigma_y)_{\alpha\beta}c_{{\bf
      k}\beta},
\end{equation}
with $\xi=\pm$, if either the strain is absent or its direction fulfills
\begin{align}
  \theta={}&\pm\frac{\pi}{4}\equiv\chi\frac{\pi}{4}.
\end{align}
Here, $c^\dagger_{{\bf k}\alpha}$ creates a HH(LH) in the spin state
$\alpha=\pm$ for $m_j=\pm 3/2$ ($m_j=\pm 1/2$).  We assume that the
Fermi wave vector $k_{\parallel F}$ does not deviate much from
rotational symmetry and thus can be replaced by its angular average
value $k_{\parallel F} \equiv \braket{k_{\parallel
    F}}_\varphi$.\cite{PhysRevB.89.161307, PhysRevB.90.115306} This
situation holds for materials close to axial symmetry, i.e.,
$\gamma_2=\gamma_3$, and a small strain amplitude $\beta$. Therefore,
we transform $\left\{k_x,k_y\right\}^\top$ into polar coordinates,
$k_{\parallel F}\{ \cos(\varphi), \sin(\varphi)\}^\top$.  Thus, if a
hole, with a spin state given by $\{1,\pm\exp(\xi
i\pi/4)\}^\top/\sqrt{2}$ and $k=k_{\parallel F}$ is injected into the
two-dimensional system (including spin-independent scattering
processes) its spin is not randomized.

For the in-plane strain, the direction condition basically requires
symmetric normal strain components $\epsilon_{xx}=\epsilon_{yy}$ and a
non-vanishing shear strain component $\epsilon_{xy}$. This situation
can be generated by $\left\langle110\right\rangle$ uniaxial
strain.\cite{:/content/aip/journal/jap/101/10/10.1063/1.2730561} We
demonstrate this explicitely in App.~\ref{sec:piezo} for an
experimental setup by use of a piezo crystal as done by Habib
\textit{et al.} in Ref.~\onlinecite{Habib2007}.

As we will see in the following, the constraint on the wave vector
${\bf k}$ of persistent spin states is crucial since it reveals that we
found no conserved spin quantity for the whole $k$-space but only for
the averaged Fermi contour. However, this constraint is not surprising
when we recall the case of persistent spin states in 2DEG. If the SO
terms are linear in the wave vector, the condition for the existence
of persistent spin states is fulfilled if the Rashba SOC coefficient
is equal to the one for the linear Dresselhaus term. In this special
case, the SO field is collinear in the whole $k$-space. If the cubic
Dresselhaus term is included, we cannot find a quantity $\Sigma$ which
commutes with the Hamiltonian $\mathcal{H}$ at every wave vector,
though. Nevertheless, if the relations resulting from
$[\Sigma,\mathcal{H}]\stackrel{!}{=}0$ are Fourier decomposed, similar to the
procedure in Ref.~\onlinecite{Knap1996}, and only the lowest harmonics
in the azimuthal angle is considered, one finds a condition for
long-lived spin states. In contrast to the case without the cubic
contribution, the found symmetry is, however, bound to an appropriate
energy.\cite{DollingerThesis2013} This can also be understood by
studying the spin relaxation rates in diffusive n-type wires with
Rashba and Dresselhaus SOC. One does not only find an additional
spin-relaxation term due to the cubic Dresselhaus but also, the linear
Dresselhaus coefficient is shifted.\cite{Kettemann:PRL98:2007} This
shift depends on the Fermi energy.

Next, as in the previous section, we consider separately the case of a HH like
ground state and the LH like ground state.
For the sake of simplicity, we apply the following replacements:
\begin{align}
  \lambda_{\text{D},\pm} ={}& n_{\text{\tiny HH/LH}} \lambda_{\text{R},\pm},\\
  \gamma_3 ={}& \Gamma \gamma_2,\\
  \beta ={}& B k_{\parallel F},\label{eq:beta}\\
  \braket{k_z^2} ={}& \left(\frac{k_{\parallel F}}{\kappa}\right)^2,\\
  \intertext{and for the HH like state additionally}
    \eta_+={}& \eta_0 \gamma_2 \braket{k_z^2} \lambda_{\text{D},+} .
\end{align}
In contrast to the
discussed effect of the cubic Dresselhaus in an 2DEG, we will find
persistent and not only long-lived spin states.
\subsection{Conserved Spin Quantity in Case of a HH Like Ground State}
Making use of the definitions above and setting $n\equiv
n_{\text{\tiny HH}}$ for simplicity, we obtain for
$\Sigma_\xi$ the following equation according to
Eq.~(\ref{condition}):
\begin{align}
  0={}&\Omega_{y,+}-\xi\Omega_{x,+}\\
   ={}&  \left(\cos(\varphi)-\xi\sin(\varphi) \right)\notag\\
  {}&\cdot\left\{ B^2\Gamma\xi \left( n+\xi \right)+\chi\left[\xi+n(\eta_0/\kappa^2-1)\right] \right.\notag\\
  {}& \phantom{a}+\left[B^2\kappa^2 n \Gamma
      +2\chi \left(1+\Gamma+n\xi\left(\Gamma-1\right)\right) \right]\notag\\
    {}&\left.\phantom{aaa}\cdot\cos(\varphi)\sin(\varphi) \right\}.
\label{cond2}
\end{align}
This equation is fulfilled independently of the polar angle $\varphi$
if the ratio between Dresselhaus and Rashba SOC strength $n$ and the
strain strength factor $B$ satisfy the relations
\begin{equation}
  n_{\xi, \chi}^{(\pm)}=\xi\frac{2\left(1+\Gamma\right)}{2\left(1-\Gamma\right)-(\xi\chi) \Gamma \kappa^2 \left(B_{\xi,\chi}^{(\pm)}\right)^2}
\label{n}
\end{equation}
and
\begin{equation}
  B_{\xi, \chi}^{(\pm)}=\sqrt{\frac{\xi\chi\left(4-\kappa^2\right)\pm \mathcal{W}}{2\Gamma\kappa^2}},
  \label{B}
\end{equation}
where
\begin{equation}
 \mathcal{W}=\sqrt{\kappa^4+8\left(1+\Gamma\right)\eta_0-8\left(1+2\Gamma\right)\kappa^2+16},
  \label{W}
\end{equation}
and $-B_{\xi, \chi}^{(\pm)}$ are also solutions. 

If the $C_k$ term in Eq.~\ref{etaHH} can be neglected (full expression can
be found in App.~\ref{kappa_domain}) real solutions for $B_{\xi, \chi}^{(\pm)}$ are only
found for $\kappa \in A_{\mathrm{sgn}(\xi\cdot\chi)}$ where 
\begin{align}
  A_+=&\left[0,2\sqrt{1+2\Gamma-2\sqrt{\left(1+\Gamma\right)(\Gamma-\eta_0/8)}}\right],\\
  A_-=&\left[2\sqrt{1+2\Gamma+2\sqrt{\left(1+\Gamma\right)(\Gamma-\eta_0/8)}},\infty\right)
\end{align} 
if $\Im (A_{\mathrm{sgn}(\xi\cdot\chi)})=0$ which usually holds as typically $\Gamma>0$ and $\eta_0<0$.

In absence of strain, that is, $\beta=0$, the fomulas above yield the
following requirements on the ratio $n$ and for the
Fermi wave vector $k_{\parallel F}$ to get $\Sigma_\xi$:
\begin{align}
  n_0{}&\equiv n=\xi \frac{1+\Gamma}{1-\Gamma},\label{n0}\\
  k_0{}&\equiv k_{\parallel
    F}=\sqrt{\frac{\eta_+}{2\gamma_2\lambda_\text{D,+}}\frac{1+\Gamma}{\Gamma}}\label{k0}.
\end{align} 
In this scenario, the SO field even vanishes at a specific value $k_0$
of the Fermi wave vector in the axially symmetric case, i.e.,
$\Gamma=1$.  For this purpose, both linear and cubic Dresselhaus
contributions are crucial.  Hence, this solution was not existent in
our previous publication Ref.~\onlinecite{PhysRevB.90.115306}.  The
axially symmetric case demands a vanishing Rashba contribution,
$\lambda_\text{R,+}/\lambda_\text{D,+}\approx 0$.  However, for the
most semiconductors $\Gamma$ ranges from 1 to 1.5. Thus, the cubic
Dresselhaus SOC strength has to outweigh the Rashba SOC strength,
i.e., $ |\lambda_{\text{D},+}| >| \lambda_{\text{R},+}|$.

\begin{figure}[t]
  \includegraphics[width=\columnwidth]{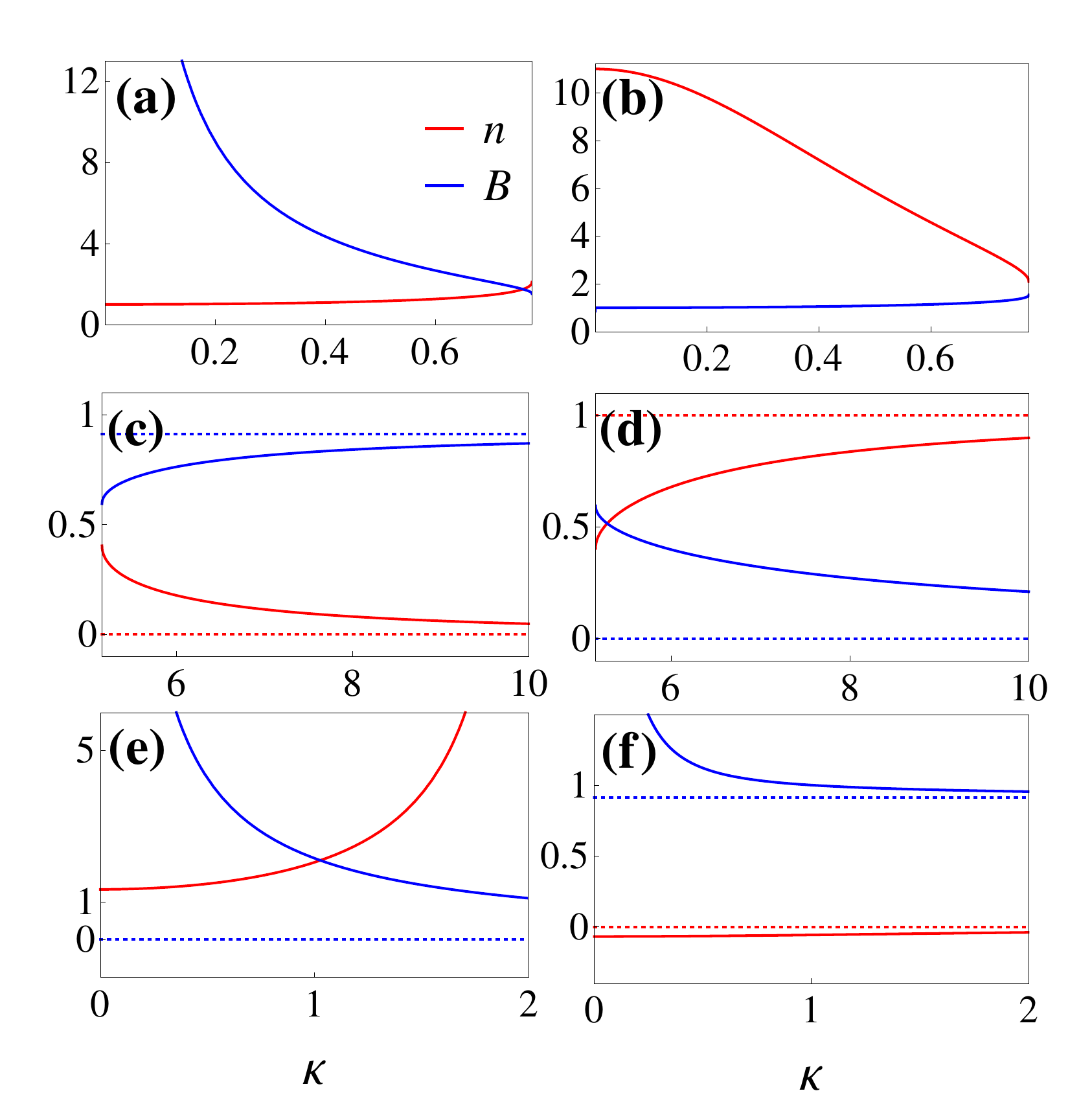}
  \caption{(Color online) Parameter configurations for $n_{\xi,
      \chi}^{(\pm)}$ and $B_{\xi, \chi}^{(\pm)}$ which yield the
    conserved spin quantity $\Sigma_\xi$ for $ \theta=\chi \pi/4$ in
    case of a HH like ((a)-(d)) and LH like ((e)-(f)) ground
    state. The ratio between the Luttinger parameter $\gamma_3$ and
    $\gamma_2$ is set to $\Gamma=1.2$ and the linear Dresselhaus
    contribution is neglected, i.e., $\eta_0\rightarrow 0$. The domain
    for $\kappa$ is $A_+$ for $\xi=\chi$ and $A_-$ else. The dashed
    lines indicate the
    according asymptotes at large width of the quantum well.
    (a) $n_{\xi, \chi}^{(+)}$ and $B_{\xi, \chi}^{(+)}$, (b) $n_{\xi,
      \chi}^{(-)}$ and $B_{\xi, \chi}^{(-)}$ for $\xi=\chi=-1$. If
    $\xi=\chi=1$ holds only the sign of $n$ is inverted. (c) $n_{\xi,
      \chi}^{(+)}$ and $B_{\xi, \chi}^{(+)}$, (d) $n_{\xi,
      \chi}^{(-)}$ and $B_{\xi, \chi}^{(-)}$ for $\xi=-\chi=1$.
    Interchanging of $\xi$ and $\chi$ only changes the sign of $n$.
    (e) $n_{\xi, \chi}^{(+)}$ and $B_{\xi,\chi}^{(+)}$ which yield the
    conserved spin quantity $\Sigma_-$ for $\xi=\chi=-1$ and (f)
    $\xi=-\chi=-1$. The case $\xi=1$ reverses only the sign of $n$.}
  \label{plot:cq}
\end{figure}

More peculiar solutions occur for $\Gamma\neq 1$ or in the presence of
strain. For certain parameter configurations, the SO field becomes
collinear on the averaged Fermi contour. As a consequence, this SU(2)
spin rotation symmetry gives rise to a persistent spin
helix.\cite{Bernevig2006} For a start, by setting $\Gamma=-1$,
$\chi=\xi=1$ we can recover the solutions presented by Sacksteder
\textit{et al.}  in Ref.~\onlinecite{PhysRevB.89.161307}.  In this
case, we obtain $n=0$, which is obvious since Dresselhaus SOC was not
considered in Ref.~\onlinecite{PhysRevB.89.161307}, and $B=1$, which
is consistent with their results.  It is remarkable that the presence
of the linear Dresselhaus term, which was not considered either by
Sacksteder \textit{et al.}, does not alter the result.  Note that
here, too, the solution for the conserved spin quantity is bound to an
averaged Fermi contour by Eq.~(\ref{eq:beta}).  Moreover, recalling
Eq.~(\ref{condition}), one finds for these particular parameters a
conserved spin quantity for every direction $\theta$, given by
\begin{align}
  \Sigma=\sin(\theta)\sigma_x-\cos(\theta)\sigma_y,
\end{align}
which generalizes the result from the previous publication.  Yet, we
emphasize that the condition $\Gamma=-1$ is rather unusual for most
materials.
Regarding the realization of persistent spin states in experiments, it
is of interest to analyze whether the constrains allow for realistic
(i.e., typical for III-V semiconductors) parameters. Thus, we present
a concrete example for the conserved quantities $\Sigma_\xi$. At this,
we assume $\gamma_3-\gamma_2>0$, Eq.~(\ref{gamma3gamma2}), and
$\gamma_i>0$ to hold and choose, as an example, $\Gamma=1.2$ for the
plots of $n_{\xi, \chi}^{(\pm)}$ and $B_{\xi, \chi}^{(\pm)}$,
Fig.~\ref{plot:cq}.  In order to draw a general picture and for
simplicity we neglected the linear Dresselhaus contribution, i.e.,
$\eta_0\rightarrow 0$.  In the range $A_+$, i.e., $\xi=\chi$ that
comprises realistic values for $\kappa$ (e.g., $\kappa=0.3$ for a
confinement width $L=100\, \text{\AA}$ and a small Fermi vector
$k_{\parallel F}=0.01\, \text{\AA}^{-1}$) solutions are displayed in
Fig.~\ref{plot:cq}(a,b). The solutions with a small $B$ value,
$B^{(-)}_{-, -}$ in Fig.~\ref{plot:cq}(b), are preferable since in
this case the deformation of the Fermi contour is small. In addition,
it is reasonable to assume that $n>1$, since the Dresselhaus
contribution is usually larger than the Rashba contribution.
The domain $A_-$ ($\xi=-\chi$) shown in Fig.~\ref{plot:cq}(c,d) is
less realisic.  Having large values of $\kappa$ implies a large width
of the quantum well and a high Fermi energy which leads to populations
in higher subbands where the model loses its validity. Also, for a
large Fermi energy the Fermi contour is strongly deformed as can be
understood by examining the term proportional to $\beta^2\gamma_3^2$
of the kinetic energy, Eq.~(\ref{Ekin_HH}). A spherical approximation
becomes inappropriate. Moreover, if a strong strain is applied to the
sample the appropriate model Hamiltonian needs to include also the
coupling to the split-off band.

For future devices like the spin field-effect-transistor it is not
only of interest to find persistent spin states. In fact, samples are
favorable where the injected particles undergo only a well defined
spin-rotation. Thereby the initial spin state, with $\bf k$ being a
good quantum number, is not necessarily an eigenstate. Here,
\textit{well defined} means that the rotation only depends on the
distance between the injection and detection position.  For n-type
systems this condition was already analyzed in
Ref.~\onlinecite{Schliemann2003}. Concerning a 2DHG as described in
this paper, a spin-conserving condition which is valid for spin states
with arbitrary wave vector cannot be found: The condition is limited
to the averaged Fermi contour. For these states we can find an
additional condition so that their precession depends only on the
distance. At this, a necessary condition is an elastic scattering from
impurities.  The corresponding effective vector field in the case
where the Eqs.~(\ref{n},\ref{B}) hold has the structure given by
\begin{align}
  \boldsymbol{\Omega}_{\text{HH}} ={}& (k_x+\xi
  k_y)\varphi^{(\pm)}(k_x,k_y)\begin{pmatrix} \xi \\ 1 \\
    0 \end{pmatrix}\Bigg|_{k_x^2+k_y^2=k_{\parallel
      F}^2}.\label{optimalVF}
\end{align}
In the case where $(k_x+\xi k_y)\varphi^{\pm}(k_x,k_y)$ depends
linearly on $k_i$ ($k_{\parallel F}$ is a constant), the mentioned spin-rotation is only distant
dependent. Here, one finds
\begin{align}
 \varphi^{(\pm)}{}&(k_x,k_y)\notag\\
= {}&\frac{k_{\parallel F}^2
   \gamma_2\lambda_{R,+}}{\kappa^2(\pm\xi\chi \mathcal{W}+4\Gamma-\kappa^2)}\Big[2 (4 \pm\xi\chi \mathcal{W}) (1 +
 \Gamma)\notag\\
{}&-(2 \pm\xi\chi \mathcal{W} + 6 \Gamma)
\kappa^2 + \kappa^4 + \xi 8 (\Gamma^2-1) \kappa^2 \frac{k_x k_y}{k_{\parallel F}^2}\Big].
\end{align}
Thus, the special case where a well defined spin rotation occurs can
only be found if $\Gamma\equiv\gamma_3/\gamma_2=\pm 1$.
\subsection{Conserved Spin Quantity in Case of a LH Like Ground State}
Analogously, it is possible to find conserved spin quantities if the
ground state is LH like.  Yet, the structure of the SO field
is more complex since it contains additional first order term due to
Dresselhaus SOC, Eq.~(\ref{omega_b}). As stated above, an [110]
uniaxial compressive strain leads to $\delta<0$ and therefore cannot
be used. It is commonly known that for in-plane biaxial tensile stress
LH ground state can be created, but in that case $\epsilon_{xy}$
vanishes\cite{:/content/aip/journal/jap/101/10/10.1063/1.2730561}.
Consequently, combined strain effects are necessary to generate the
required condition. Nonetheless, we stress that we do not demand a
strong in-plane strain amplitude for identifying a conserved spin
quantity. In fact, an appropriate tensor component $\epsilon_{zz}$ is
necessary. This component is encapsulated in the splitting
\begin{align}
  \Delta_\text{h1,ln} ={}&
  \frac{\hbar^2}{2m_0}\braket{k_z^2}\left[\left(n^2-1\right)\gamma_1+2\left(n^2+1\right)\gamma_2\right]-\delta,
 \end{align}
 and thus in the SOC strength. For simplicity, we set $n\equiv
 n_{\text{LH}}$. Hence, the calculation of the conserved quantity is
 the same as before and valid as long as the deformation of the Fermi
 contour is not excessively strong. In this case, we find for the
 parameters $n_{\xi, \chi}^{(\pm)}$ and $ B_{\xi, \chi}^{(\pm)}$ the
 relations
\begin{align}
  n_{\xi, \chi}^{(\pm)}=\frac{6\left(\Gamma-1\right)}{3 \chi \Gamma
    \left(B_{\xi, \chi}^{(\pm)}\right)^2
    \kappa^2-2\xi\left(3\left(\Gamma+1\right)+2 \mathcal{Q} \right)}
\label{nLH}
\end{align}
and
\begin{align}
  B_{\xi,
    \chi}^{(\pm)}=\sqrt{\frac{\xi\chi\left(4\mathcal{Q}-3\kappa^2+12\right)\pm
      \mathcal{P}}{6 \Gamma \kappa^2}},
  \label{BLH}
\end{align}
where we defined
\begin{align}
  \mathcal{P}={}&\sqrt{16\left(\mathcal{Q}+3\Gamma\right)^2+24\kappa^2\left(\mathcal{Q}+6\Gamma-3\right)+9\kappa^4},\\
  \mathcal{Q}={}&\frac{\Delta_{\text{l1,h1}}}{\left|(\Delta_{\text{l1,h1}}\big|_{\delta=0})\right|}.
\end{align}
The parameters $n_{\xi, \chi}^{(\pm)}$ and $B_{\xi, \chi}^{(\pm)}$ are
plotted in Fig.~\ref{plot:cq}(e,f) for $\Gamma=1.2$ and
$\mathcal{Q}=1$ which is equivalent to an energy shift
$\delta=2\left|(\Delta_{\text{l1,h1}}\big|_{\delta=0})\right|$. We
only find real solutions for $B_{\xi, \chi}^{(+)}$ and $n_{\xi,
  \chi}^{(+)}$.  In contrast to the HH like ground state for a
realistic system with $\Gamma>1$ we do not find a conserved spin
quantity if strain is absent.

In the last part of this section we apply the insights on the
conserved spin quantity to a prominent semiconductor.
\subsection{Example: p-doped InSb}\label{sec:example}
We choose p-doped InSb as an example to contrast the strained case
yielding a conserved spin quantity with the one of a strainless
sample. We assume a confinement in [001] direction with a depth of
$L=\SI{100}{\AA}$. To guarantee a low filling we set $k_{\parallel
  F}=\SI{0.01}{\AA^{-1}}$. The used parameters here are listed in
App.~\ref{sec:bia_parameters}. Further, we assume the additional
splitting due to strain between HH and LH like subbands to vanish,
$\delta=0$. Choosing an external electric field of
$\mathcal{E}_z=\SI{1.6}{kV/cm}$ and a [110] tensile strain direction
($\epsilon_{xy}>0$), i.e., $\xi=\chi=1$, allows for a persistent spin
polarization in [110] direction. Since we can assume a HH like ground
state, we apply Eq.~(\ref{n}) and (\ref{B}) and obtain the parameter
for the in-plane strain strength to be $B_{1,1}^{(-)}=0.74$ and the
corresponding ratio between Dresselhaus and Rashba SOC strength to be
$n_{1,1}^{(-)}=-20.7$.
\begin{figure}[htbp]
  \includegraphics[width=1.05\columnwidth]{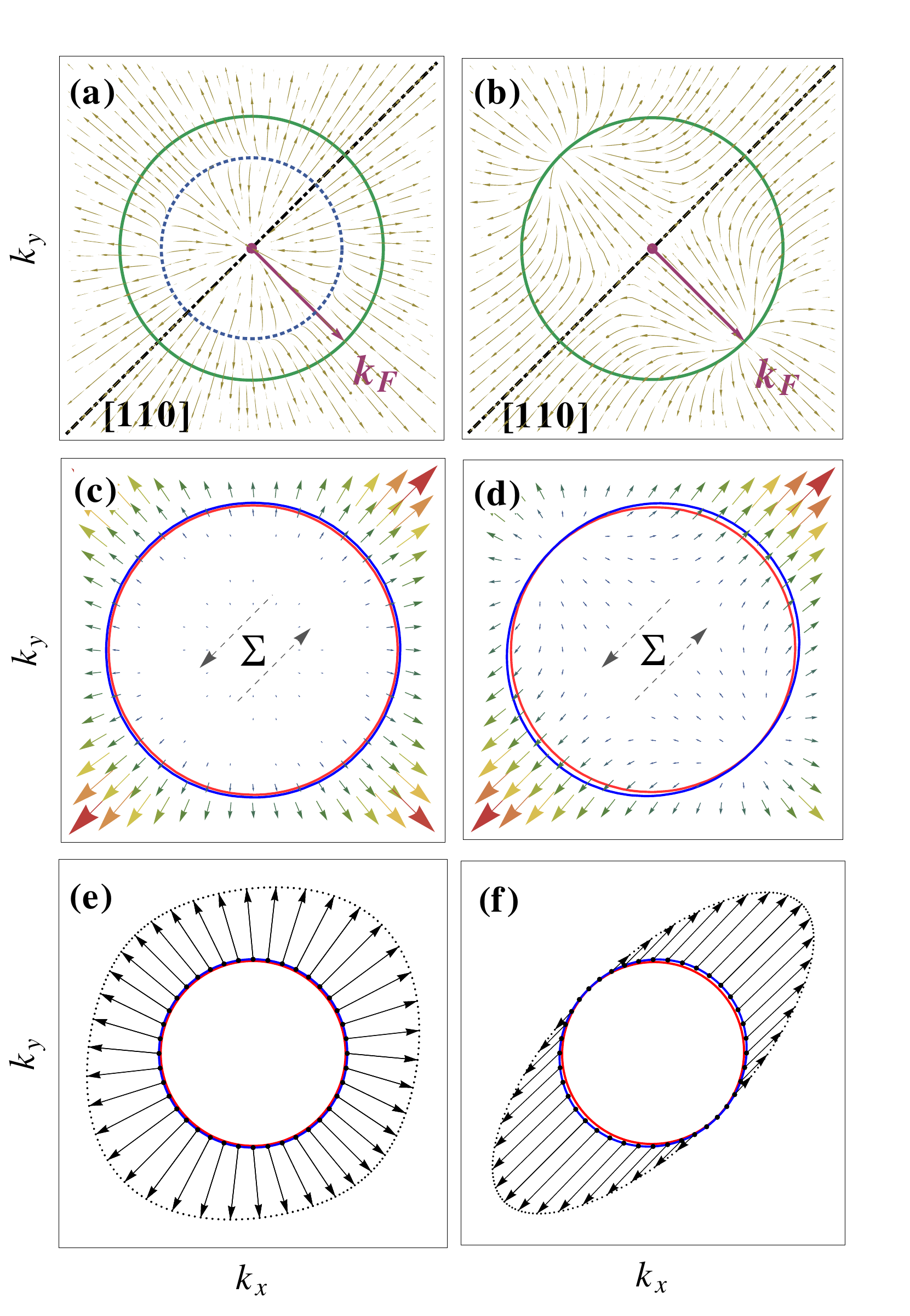}
  \caption{(Color online) Comparison of the SO field without applied
    strain (left column) and the case where strain gives rise to a
    conserved spin quantity $\Sigma$ in case of a HH like ground
    state. In the presented case the SOC strengths are
    $\lambda_\text{D,+}=\SI{-42.5}{eV\AA^3}$, $\lambda_\text{R,+} =
    \SI{2.05}{eV\AA^3}$ and $\eta_+=\SI{-34.1e-3}{eV\AA}$.  (a) and
    (b) Stream plot of the effective SO vector field. The green circle
    indicates the axially symmetric Fermi contour. The blue dotted
    circle corresponds to $k_0=\SI{6.9e-3} {\AA^{-1}}$ where the field
    vanishes approximately.  (c) and (d) Vector field and Fermi
    contours. The gray arrows indicate the spin polarization.  (e) and
    (f) Detailed picture of the SO field that operates at the outer
    Fermi contour.}
  \label{plot:SOF}
\end{figure}
\begin{figure}[htbp]
  \includegraphics[width=.7\columnwidth]{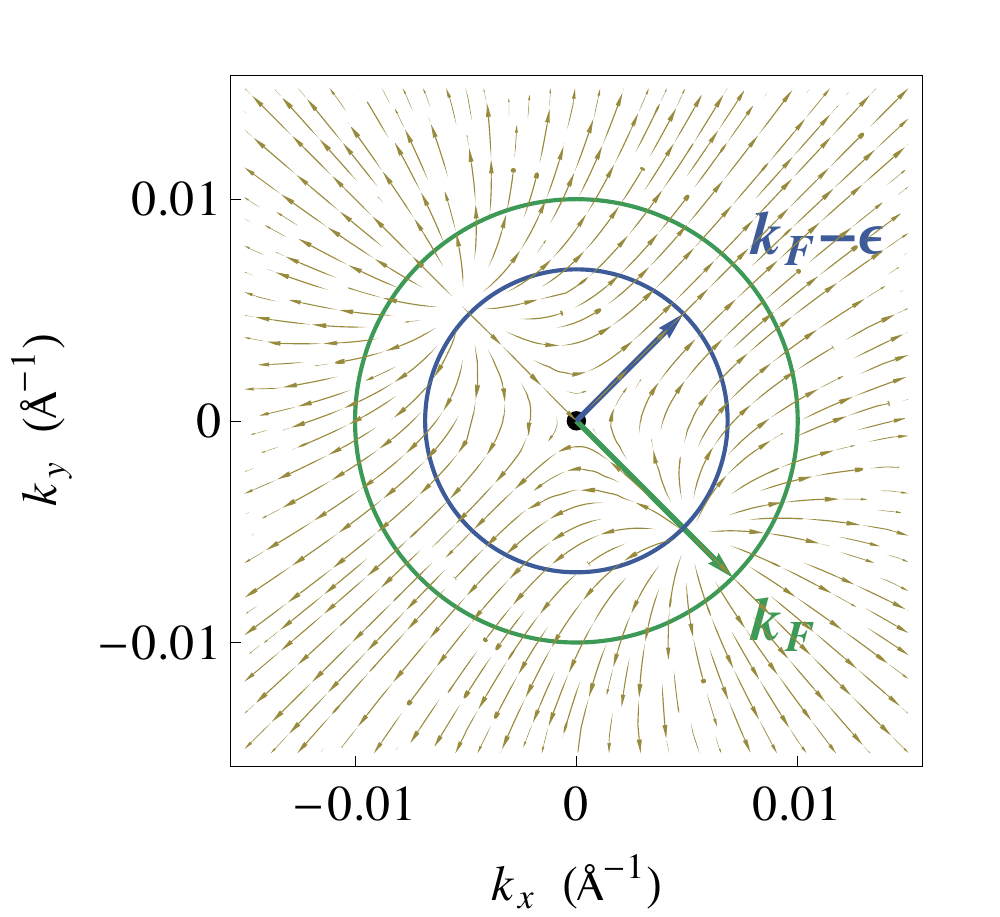}
  \caption{(Color online) Stream plot of the effective SO vector field
    without linear Dresselhaus contribution, i.e., $\eta_+\rightarrow
    0$, in case of a HH like ground state. The strain-induced sources
    of the field move to a slightly lower Fermi wave vector by
    $\epsilon=\SI{3.1e-3}{\AA^{-1}}$ where the spin-preserving
    symmetry is recreated.}
  \label{plot:SOFshift}
\end{figure}
In Fig.~\ref{plot:SOF}(d) the resulting effective SO field is plotted
and compared to the case where the [110] stress is absent,
Fig.~\ref{plot:SOF}(c).  A stream plot, Fig.~\ref{plot:SOF}(a,b) shows
that without strain, Fig.~\ref{plot:SOF}(a), the vector field vanishes
approximately at $k_0=\SI{6.9e-3} {\AA^{-1}}$ which is illustrated by
the blue dotted circle. We see that even though the condition
Eq.~(\ref{n0}) on $n_0$ for the spin-preserving symmetry in the
strainless case is not perfectly fulfilled, i.e.,
$n_{1,1}^{(-)}>n_0=-29.6$, the location where the field disappears is
still well described by $k_0$ in Eq.~(\ref{k0}). Additionally, there
is one source in the vector field at $k_\parallel=0$. Including
strain, Fig.~\ref{plot:SOF}(b), gives rise to two additional sources
that are centered at the crossing of the Fermi contour and the
$[1\overline{1}0]$ axis. This can be understood by considering the
factor $(k_x+\xi k_y)$ in Eq.~(\ref{optimalVF}). At these two sources
vector field components are suppressed which are not collinear with
the $[110]$ direction. The Fermi contours split due to the
SOC. Without strain, they are only slightly deformed as consequence of
the band warping. If strain is present, the deviation of rotational
symmetry of the contours is enhanced. The deformation is most intense
in the $[110]$ and $[1\overline{1}0]$ direction. To guide the viewer's
eye, we give in Fig.~\ref{plot:SOF}(e,f) a detailed picture of the SO
field acting on the outer Fermi contour. In the case of strain the
vectors lie parallel to the [110] direction. As discussed above, the
plots also show that the preserved spin quantity is limited to the
averaged Fermi contour. The vector field regions which are
noncollinear are strongly suppressed, though. This leads to a
reduction of spin relaxation even in the case of a general spin state
injected into the 2DHG.

\subsubsection*{Influence of linear Dresselhaus terms}

Moreover, we want to emphasize that in the chosen parameter regime the
effect of the linear Dresselhaus contribution is only small.
Fig.~\ref{plot:SOFshift} shows how the field modifies if the linear
Dresselhaus term proportional to $\eta_+$ is neglected.  The
strain-induced additional field sources move to a marginally lower
Fermi wave vector by $\epsilon=\SI{3.1e-3}{\AA^{-1}}$ where the
conserved spin quantity is reobtained.  To first order in $\eta_+$ the
shift $\epsilon$ can be generally estimated by
\begin{align}
  \epsilon ={}& \frac{\eta_+}{\sqrt{2}\beta\gamma_3}\notag\\
  {}&\times\phantom{=}\left\lbrace (\lambda_\text{D,+}+\lambda_\text{R,+})
    \left[\frac{\beta^2\lambda_\text{D,+}}{\braket{k_z^2}} +
      2(\lambda_\text{D,+}+\lambda_\text{R,+})\right]\right\rbrace^{-1/2}.
\end{align}
We stress that the influence of the linear BIA terms becomes even
smaller for increasing Fermi wave vector $k_F$ and in other materials
such as GaAs is less significant.

\section{Summary}
Summarizing, we identified conserved spin quantities in a
(001)-confined two-dimensional hole gas in semiconductors with
zincblende structure. Thereby, we derived the dominant contribution to
the SO field due to Rashba SOC directly from an electric field
$\mathcal{E}_z$, which was missing in our previous publication,
Ref.~\onlinecite{PhysRevB.90.115306}. The significant effect due to
Rashba SOC is only controlled by the subband gaps and not, as in the case
of Dresselhaus SOC, by the conduction band gap.  In view of recent
publications, we also included the effect of linear Dresselhaus SOC
terms whose significance was pointed out in
Refs.~\onlinecite{PhysRevB.89.075430, PhysRevLett.104.066405} to be
underestimated.  The proper determination of the SO field
enabled us to conclude that there are two possiblities for long-lived
spin states.  In respect of an unstrained sample such states exist
only for heavy holes. It requires a certain ratio of cubic Rashba and
Dresselhaus SOC strength defined solely by the Luttinger parameters
$\gamma_2$ and $\gamma_3$.  Other spin-preserving symmetries occur in
presence of strain for both a HH like and LH like ground state.  Here, a
non-vanishing [110] shear strain component $\epsilon_{xy}$ and a
symmetric in-plane normal strain $\epsilon_{xx}=\epsilon_{yy}$ are
essential.  We have recovered the conserved spin quantity presented in
Ref.~\onlinecite{PhysRevB.89.161307} for the special case where
$\gamma_2/\gamma_3=-1$.  In all circumstances, owing to the presence
of both linear and cubic terms due to SOC the persistent spin states are bound to
a Fermi contour.  We have also demonstrated that only for
this case and for $\gamma_2/\gamma_3=1$ one finds a spin rotation of a
spin on the averaged Fermi contour which only dependents on the
distance between the injection and detection position. Moreover, we
have shown that for the existence of a conserved spin quantity in
semiconductors which are accessible for experiments (e.g., systems
with $\gamma_2/\gamma_3\approx 1$) the interplay between Dresselhaus
SOC, Rashba SOC and possibly strain is crucial.

In this way, shear strain has turned out to be a key component for an
efficient manipulation of spin lifetime in 2D hole systems of
zincblende structure.

\begin{acknowledgments}
  The authors thank Tobias Dollinger, Andreas Scholz, Klaus Richter,
  Roland Winkler, Mikhail Glazov and Eugene Sherman for fruitful
  discussions. This work was supported by Deutsche
  Forschungsgemeinschaft via Grant No.~SFB 689.
\end{acknowledgments}
\appendix \section{Luttinger Parameter
  Relation}\label{luttinger_param_relation} 
Here, we shortly focus on the Luttinger parameters $\gamma_2$ and
$\gamma_3$. The warping of the valence band is directly proportional
to their difference. Comparing with experimental results, for most
semiconductors one finds the parameter $\gamma_3$ to be larger than
$\gamma_2$.\cite{:/content/aip/journal/jap/89/11/10.1063/1.1368156,
  YuCardonaBook, winklerbook} This becomes clear when describing the
system using ${\bf k\cdot p}$ method for band structure calculations,
which yields the relation\footnote{Eq.~(\ref{gamma3gamma2}) is valid
  if the reduced Luttinger parameters $\gamma_2^\prime$ and
  $\gamma_3^\prime$ in the applied model vanish\cite{Mayer1991}.}
\begin{align}
  \gamma_3-\gamma_2 = \frac{2}{3m_0 E_0^\prime}Q^2.\label{gamma3gamma2}
\end{align}
Hereby, we follow the notation and phase conventions of
Ref.~\onlinecite{YuCardonaBook} where $Q\in \mathbb{R}$ and $iQ$ being
a momentum matrix element between the $\Gamma_{8v}$\footnote{Following
  the Koster notation\cite{Koster1957173}, one has for the tetrahedral
  point group $T_d$ three double group representations given by the
  two two-dimensional representations $\Gamma_6$, $\Gamma_7$ and the
  four dimensional one $\Gamma_8$.} valence and the $\Gamma_{7c}$ and
$\Gamma_{8c}$ conduction band states. The energy
separation between the conduction band $\Gamma_{7c}$ and the $j=3/2$
valence bands is denoted as $E_0^\prime>0$.\cite{YuCardonaBook}
\section{Applied Approximations}
\subsection{L\"owdin's Partitioning}\label{secloewdin}
In this paper we start with a Hamiltonian which describes HH and LH in
the bulk. At the $\Gamma$ point both types are degenerate. Imposing a
confinement on the system reduces it to a quasi 2D system, generating
HH like and LH like subbands. The simplification, which allows for
further analytical studies, is now to focus on the subspace spanned by
either the set $\{\ket{j=3/2,m_j=\pm1/2}\ket{n=1}\}$ or
$\{\ket{j=3/2,m_j=\pm3/2}\ket{n=1}\}$, only, where $n$ is the subband
index. Let us call this subset $A$. Due to the confinement in growth
direction and strain, this subset is well separated in energy at the
$\Gamma$ point from all other subbands (except the particular case
where strain is exactly reversing the energy splitting due to the
confinement between the lowest band and the subsequent one at the
$\Gamma$ point). Assuming $A$ having HH character, let us call the
subset spanned by $\{\ket{j=3/2,m_j=\pm3/2}\ket{(n>1)},$
$\ket{j=3/2,m_j=\pm1/2}\ket{n}\}$ subset $B$. To end up with an
effective model which focuses on the lowest HH(LH) like subbands, one
treats the effect of subset $B$ on the subset $A$ in a perturbative
manner. Here, one has to distinguish between degenerate and
non-degenerate perturbation theory since $A$ and $B$ may contain exact
or approximate degeneracies. Applying the quasi-degenerate
perturbation theory, also called L\"owdin's partitioning, avoids this
tricky task. It is described in great detail in the book by Bir and
Pikus\cite{BirPikusSymmetryStrain1974} or by
Winkler\cite{winklerbook}. With the subspace decomposition of our
problem in mind, we give in the following a short formal description
of the perturbation procedure. Assume that $\mathcal{H}$ can be
expressed as a sum of a Hamiltonian $\mathcal{H}^0$ with known
eigenvalues $E_n$ and eigenfunctions $\ket{\psi_n}$ and $\mathcal{H'}$
which is treated as a perturbation. Further, assume $\mathcal{H'}$
being a sum of a block diagonal matrix $\mathcal{H}^1$ with subsets
$A$ and $B$ and $\mathcal{H}^2$ describing the coupling of both
subsystems,
\begin{align}
  \mathcal{H}=\mathcal{H}^0+\mathcal{H'}=\mathcal{H}^0+\mathcal{H}^1+\mathcal{H}^2.
\end{align}
In accordance with Ref.~\onlinecite{winklerbook,
  BirPikusSymmetryStrain1974}, we define the indices $m, m', m''$
corresponding to the states in set $A$, the indices $l,l',l''$ to
states in set $B$ and the matrix elements between them as
\begin{align}
\mathcal{H}'_{ij}=\braket{\psi_i|\mathcal{H}'|\psi_j}.
\end{align}
Now one can find a non-block-diagonal anti-Hermitian matrix $S$ which
allows for a transformation of $\mathcal{H}$ yielding a block diagonal Hamiltonian
$\widetilde{\mathcal{H}}=e^{-S}\mathcal{H} e^{S}$. $S$ and thus $\mathcal{H}$ can be found
from a successive approximation. The first terms
\begin{align}
\widetilde{\mathcal{H}}={}&\mathcal{H}^{(0)}+\mathcal{H}^{(1)}+\mathcal{H}^{(2)}+\mathcal{H}^{(3)}+\ldots
\label{perturbationformula}
\end{align}
are given by
\allowdisplaybreaks
\begin{align}
\mathcal{H}^{(0)}_{m m'}={}& \mathcal{H}^0_{m m'}\, ,& \label{perturbationformula0}\\
\mathcal{H}^{(1)}_{m m'}={}& \mathcal{H}'_{m m'}\, ,\label{perturbationformula1}\\
\mathcal{H}^{(2)}_{m m'}={}& \frac{1}{2}\sum_{l}\mathcal{H}'_{m l}\mathcal{H}'_{l m'}\left[\frac{1}{E_m-E_l}+\frac{1}{E_{m'}-E_l}\right],&\label{perturbationformula2}\\
\mathcal{H}^{(3)}_{m m'}={}&-\frac{1}{2}\sum_{l, m''}
\left[
\frac{\mathcal{H}'_{m l}\mathcal{H}'_{l m''}\mathcal{H}'_{m'' m'}}
{(E_{m'}-E_l)(E_{m''}-E_l)}\right.\notag\\
{}&\phantom{-\frac{1}{2}\sum_{l, m''}}+\left.\frac{\mathcal{H}'_{m m''}\mathcal{H}'_{m'' l}\mathcal{H}'_{l m'}}
{(E_{m}-E_l)(E_{m''}-E_l)}
\right]&\notag\\
&+\frac{1}{2}\sum_{l, l'}
\mathcal{H}'_{m l}\mathcal{H}'_{l l'}\mathcal{H}'_{l' m'}
\left[
\frac{1}
{(E_{m}-E_l)(E_{m}-E_{l'})}\right.\notag\\
{}&\phantom{-\frac{1}{2}\sum_{l, m''}}\left.+\frac{1}
{(E_{m'}-E_l)(E_{m'}-E_{l'})}
\right].&\label{perturbationformula3}
\end{align}
Note that each of the subsets $A$ and $B$ may be degenerate but it is
crucial that the subsets are chosen to be separated in energy,
i.e., $E_m \neq E_l$.
\subsection{Dominant Invariants for the Cubic BIA Spin Splitting in
  Bulk Semiconductors}\label{dominantInvariants}
\begin{figure}
  \centering
  \includegraphics[width=\linewidth]{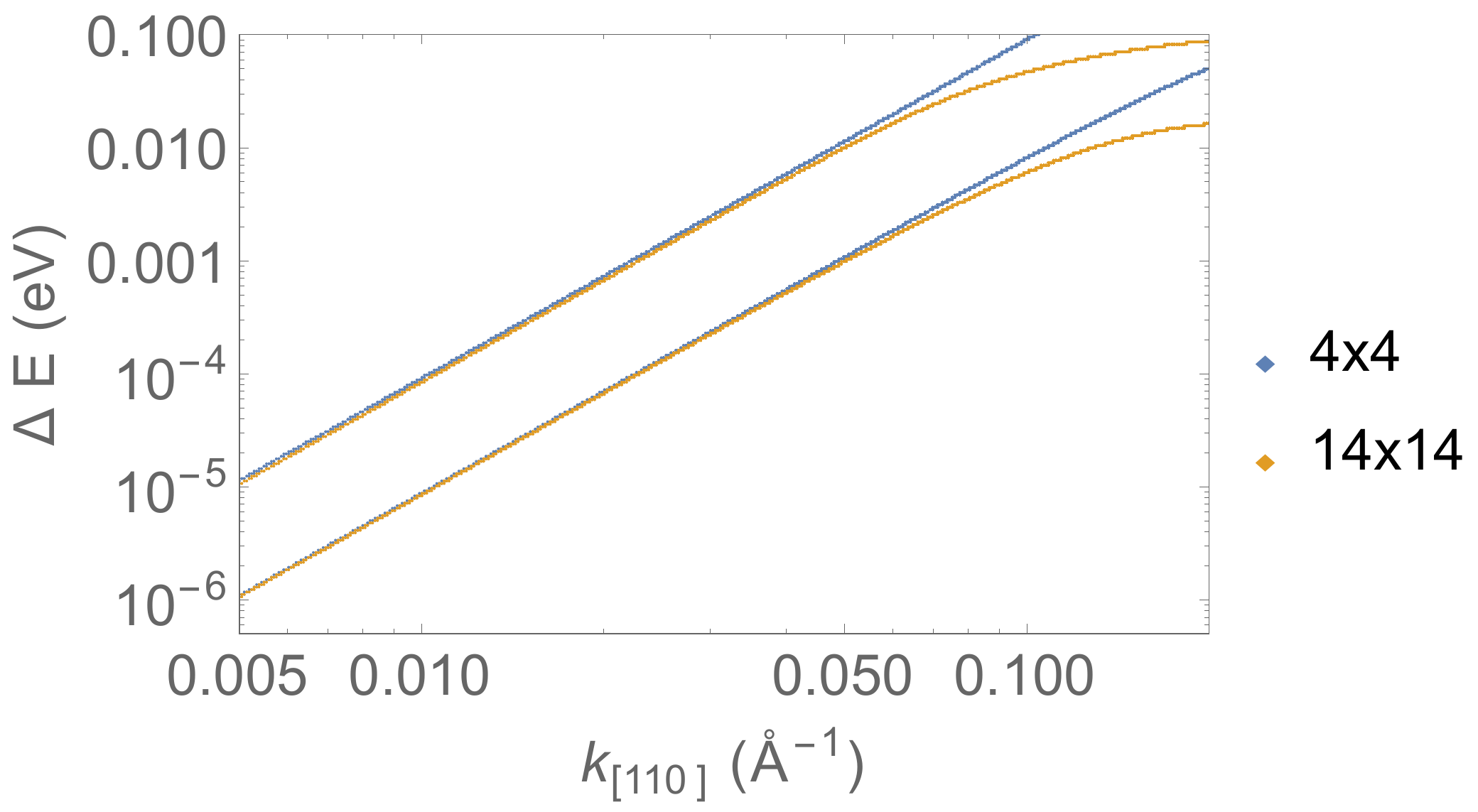}
  \caption{(Color online) Absolute value $|\Delta E|$ of the BIA spin splitting of LH
    and HH states in GaAs for ${\bf k}||[110]$ in the bulk system. The
    results are obtained by means of diagonalization of the full $14
    \times 14$ matrix of the extended Kane model (orange) and the
    effective $4 \times 4$ model used in this paper (blue), following
    Ref.~\onlinecite {winklerbook}. Here, contributions from remote
    bands to the extended Kane model are excluded.}
  \label{fig:comparison_extended_vs_4x4}
\end{figure}

Our starting point is the extended Kane model excluding contributions
from remote bands, i.e., $C_k=0$. The remaining invariants for the $\Gamma_{8v}$ band block in
this model,
which give rise to cubic BIA spin splitting,
are given by Eq.~(\ref{bia}).
In order to determine the coefficients, L\"owdin's partitioning in the
energy gaps at the $\Gamma$ point in the extended Kane model can be
applied. In Ref.~\onlinecite{winklerbook} [Tab.6.3.], the coefficients
in third order perturbation theory have been listed for several
compounds. It reveals that in the bulk compared to $b_{41}^{8v8v}$, the terms
proportional to $b_{42}^{8v8v}$, $b_{51}^{8v8v}$ and $b_{52}^{8v8v}$
can be neglected. As an example, we compare the extended Kane
model with the effective $4 \times 4$ model used in this paper by
calculating the absolute value $|\Delta E|$. The latter is the BIA
spin splitting calculated for both LH and HH states in GaAs for
${\bf k}||[110]$ in the bulk system. The result is plotted in
Fig.~\ref{fig:comparison_extended_vs_4x4} and shows very good
agreement between both models. Deviations are only present at large
$k$ values.
Note, however, that for a certain choice of parameters for the ${\bf
  k}\cdot {\bf p}$ model higher order corrections to the coefficients
can be significant, as recently shown in
Ref.~\onlinecite{PhysRevB.89.075430}. For a comparison of ${\mathcal
  H}_{\text{BIA}}$ with the terms used in
Ref.~\onlinecite{Rashba1988175} or
Ref.~\onlinecite{PhysRevB.89.075430} it is useful to recast
Eq.~(\ref{bia}) for $C_k=0$ in the form
\begin{align}
  {\mathcal H}_{\text{BIA}}={}& b_{41}^{8v8v}({\bf J}\cdot{\boldsymbol {\kappa}}) +
  b_{42}^{8v8v}\sum_\alpha J_\alpha^3 \kappa_\alpha+(b_{52}^{8v8v}-b_{51}^{8v8v})\nonumber\\
  {}&\times\sum_\alpha V_\alpha k_\alpha\left(
    k_\alpha^2 + \frac{1}{(b_{52}^{8v8v}/b_{51}^{8v8v})-1 }k^2
  \right),
\end{align}
with $V_x=\{J_x,J_y^2-J_z^2\}$, $J_\alpha$ and $\kappa_x=k_x(
k_y^2-k_z^2 )$ (and corresponding terms).
The relation between the coefficients used in
Ref.~\onlinecite{Rashba1988175} and the one in Eq.~(\ref{bia}) are
thus given by
\begin{align}
  b_{41}^{8v8v} = \frac{i}{6}P P^\prime Q
  \frac{1}{E_0}\left(\frac{13}{E^\prime_0} - \frac{5}{E^\prime_0+\Delta_0^\prime} \right) \hat ={}&\, \alpha_v+\frac{13}{8}\delta\alpha_v\label{b41},\\
  b_{42}^{8v8v} = \frac{2i}{3}P P^\prime Q
  \frac{1}{E_0}\left(\frac{1}{E^\prime_0} - \frac{1}{E^\prime_0+\Delta_0^\prime} \right) \hat ={}& -\frac{1}{2}\delta\alpha_v\label{b42},\\
  b_{51}^{8v8v} = \frac{2i}{9}P P^\prime Q \frac{1}{E_0} \left(
    \frac{1}{E^\prime_0+\Delta_0^\prime} -
    \frac{1}{E_0^\prime} \right) \hat = {}& -\frac{1}{6}\delta\alpha_v\\
  b_{52}^{8v8v} = \frac{4i}{9}P P^\prime Q \frac{1}{E_0} \left(
    \frac{1}{E_0^\prime} - \frac{1}{E^\prime_0+\Delta_0^\prime}
  \right) \hat ={}& \frac{1}{3}\delta\alpha_v,\label{b52}
\end{align}
with
\begin{align}
  P ={}& \frac{\hbar}{m_0}\braket{S|p_x|X},\\
  P^\prime ={}& \frac{\hbar}{m_0}\braket{S|p_x|X^\prime},\\
  Q ={}& \frac{\hbar}{m_0}\braket{X|p_y|Z^\prime}
\end{align}
and
\begin{align}
  \Delta_0 ={}& - \frac{3i\hbar}{4m_0^2 c^2} \braket{X|[(\nabla
    V_0)\times {\bf p}]_y|Z},\\
  \Delta_0^\prime ={}& - \frac{3i\hbar}{4m_0^2 c^2} \braket{X^\prime|[(\nabla
    V_0)\times {\bf p}]_y|Z^\prime},\\
  \Delta^- ={}& - \frac{3i\hbar}{4m_0^2 c^2} \braket{X|[(\nabla
    V_0)\times {\bf p}]_y|Z^\prime},
\end{align}
with  $V_0$ the Coulomb potential of
the atomic core, $(X,Y,Z)$ the topmost bonding p-like valence band
states and the antibonding s-like (S) and p-like
$(X^\prime,Y^\prime,Z^\prime)$ states in the lowest conduction band.

Notice that there is a difference in the application of perturbation theory to get
the prefactors: According to Eq.~[19a] in
Ref.~\onlinecite{PikusTitkov23.1988}, e.g., one finds for
$\Delta_0^\prime \approx 0$ the term, which corresponds to
$b_{42}^{8v8v}$, to be
\begin{align}
b_{42}^{8v8v} \sim P P^\prime Q
\frac{1}{E_0(E_0+E_0^\prime)}\label{b42Sherman}.
\end{align}
In contrast, Eq.(\ref{b42}) vanishes if
$\Delta_0^\prime$ can be neglected.
\subsection{LH like Valence Band Ground State: Mixing of the Electric
  Field and Dresselhaus Term}\label{sec:lh-like-valence}
Assuming a LH like valence band ground state and applying L\"owdin
perturbation to third order, terms appear in third order which couple
the electric field $\mathcal{E}_z$ with the Dresselhaus term
proportional to $b_{41}^{8v8v}$, 
\allowdisplaybreaks
\begin{align}
  \Omega_{\text{mix},x}^{(3)} ={}& -\frac{256 e\mathcal{E}_z b_{41}^{8v8v}}{27 \Delta_{\text{l1,l2}}^2}\left(2b_{41}^{8v8v}k_y (\K)+e\mathcal{E}_z k_x\right)\notag,\\
  \Omega_{\text{mix},y}^{(3)} ={}& -\frac{256 e\mathcal{E}_z b_{41}^{8v8v}}{27 \Delta_{\text{l1,l1}}^2}\left(2b_{41}^{8v8v}k_y (\K)-e\mathcal{E}_z k_x\right)\notag,\\
  \Omega_{\text{mix},z}^{(3)} ={}& 0.
\end{align}
%
It is negligible if compared to those
proportional to $\lambda_{\text{R},\pm}$, $\lambda_{\text{D},\pm}$ or $\eta_-$,
Eqs.~(\ref{etaLH})-(\ref{lambdaRM}), since we assume the SOC to be a
small correction. Therefore, we neglect these terms in the calculation
of the conserved spin quantity.
\section{Domain}
\label{kappa_domain}
If the $C_k$ term in Eq.~\ref{etaHH} cannot  be neglected the real solutions for $B_{\xi, \chi}^{(\pm)}$ are only
found for $\kappa \in A_{\mathrm{sgn}(\xi\cdot\chi)}$ where
\begin{align}
  A_+ ={}& \left[0,2\sqrt{1+2\Gamma+\frac{\sqrt{3}}{2}\frac{C_k}{\tilde\lambda_\text{D}}(1+\Gamma)-2\sqrt{\left(1+\Gamma\right)\Lambda}}\right],\\
  A_- ={}& \left[2\sqrt{1+2\Gamma+\frac{\sqrt{3}}{2}\frac{C_k}{\tilde\lambda_\text{D}}(1+\Gamma)+2\sqrt{\left(1+\Gamma\right)\Lambda}},\infty\right)
\end{align}
\vspace*{-0.5cm}
\begin{align}
  \text{with } \Lambda ={}&
  \frac{1}{2\tilde\lambda_\text{D}^2}(6C_k^2(1+\Gamma)+8\sqrt{3}C_k\tilde\lambda_\text{D}(1+2\Gamma)\\
  {}&+\tilde\lambda_\text{D}(3k_{\parallel F}^2(b_{42}^{8v8v}+b_{51}^{8v8v})+32\Gamma\tilde\lambda_\text{D}))\\
  \text{and }\tilde\lambda_\text{D}={}& \gamma_2 k_{\parallel
    F}^2\lambda_{\text{D},+}.
\end{align}
\section{Strain}
\subsection{Deformation Potentials}\label{sec:deformation_potentials}
The deformation potentials $a$, $b$ and $d$ can be defined in
different ways and there are several of them in the literature. We
list some relations between different definitions in
Tab.~\ref{deform}.
\begin{table}[h]
\begin{tabularx}{\columnwidth}{XXl}
  \hline\hline
  Eq.~(\ref{strain}), Ref.~\onlinecite{PhysRevB.89.161307} & Ref.~\onlinecite{winklerbook} & Ref.~\onlinecite{BirPikusSymmetryStrain1974},\onlinecite{:/content/aip/journal/jap/101/10/10.1063/1.2730561},\onlinecite{YSunThompsonNishidaBook}\footnote{The sign of $a$ has to be inverted.},\onlinecite{YuCardonaBook}\footnote{The sign of $b$ and $d$ has to be inverted.} \\
  \hline
  $a$ & $D_d-\frac{5}{6}D_u$ & $a+\frac{5}{4}b$ \\
  $b$ & $\frac{2}{3}D_u$ & $-b$  \\
  $d$ & $\frac{2}{3}D_u'$ & $-d/\sqrt{3}$  \\
  \hline\hline
\end{tabularx}
\caption{Relations between different conventions for the deformation
  potentials for the $\Gamma_{8v}$ valence band.}
\label{deform}
\end{table}
\subsection{Uniaxial Strain via Piezo Crystals}\label{sec:piezo}
Experimentally an unaxial strain can be conveniently implemented by
the application of a piezo crystal since the deformation is tunable.
In this setup, as done by Habib \textit{et al.} in
Ref.~\onlinecite{Habib2007}, the sample is fixed at one side of the
piezo crystal where we align the poling direction of the piezo with
the [110] direction.  Depending on the polarity of the applied
voltage, the piezo crystal extends (shrinks) along its poling
direction and simultaneously shrinks (extends) perpendicular to it.
Assuming the deformation being completely transmitted to the sample,
we can relate the strain coefficients of the sample, where the
principal axes correspond to the three $\left\langle 100\right\rangle$ axes, to the strain
coefficients of the piezo, which can be directly measured.  We define
the transfered strain parallel to the poling and perpendicular to it
as $\epsilon^\prime_\parallel$ and $\epsilon^\prime_\perp$.  Since due to
\textit{Hooke's law} an in-plane strain generates also a finite
out-of-plane component, the strain tensor of the sample becomes
\begin{align}
  \boldsymbol{\epsilon}=\frac{1}{2}
  \begin{pmatrix}
    \epsilon^\prime_\parallel+\epsilon^\prime_\perp  & \epsilon^\prime_\parallel-\epsilon^\prime_\perp & 0\\
    \epsilon^\prime_\parallel-\epsilon^\prime_\perp  & \epsilon^\prime_\parallel+\epsilon^\prime_\perp & 0\\
    0 & 0 & -\frac{2C_{12}}{C_{11}}\left(\epsilon^\prime_\parallel+\epsilon^\prime_\perp\right)
  \end{pmatrix},\label{piezo}
\end{align}
where $C_{12}$ and $C_{11}$ are stiffness tensor components depending
on the sample's material.  It becomes clear that in this situation the
in-plane normal strain is symmetric, i.e.,
$\epsilon_{xx}=\epsilon_{yy}$, and the shear strain component
$\epsilon_{xy}\neq 0$ as $\epsilon^\prime_\parallel$ and $\epsilon^\prime_\perp$ have
opposite sign, which corresponds to the situation demanded in
Sec.~\ref{sec:conserved}.
\section{BIA Parameters}\label{sec:bia_parameters}
In Tab.~\ref{BIA_coeff} we list the coefficients for the invariants
appearing in the BIA Hamiltonian, Eq.~(\ref{bia}), for some common
compounds (all values in eV\AA$^3$, except for $C_k$ in
eV\AA)\cite{winklerbook}.
\begin{table}[h]
  \begin{tabularx}{\columnwidth}{XXXl}
    \hline\hline
                    &   GaAs  & AlAs   & InSb    \\ \hline
            $C_k$   & -0.0034 & 0.0020 & -0.0082 \\
    $b_{41}^{8v8v}$   & -81.93  & -33.51 & -934.8  \\
    $b_{42}^{8v8v}$   & 1.47    & 0.526  & 41.73   \\
    $b_{51}^{8v8v}$   & 0.49    & 0.175  & 13.91   \\
    $b_{52}^{8v8v}$   & -0.98   & -0.35  & -27.82  \\
    \hline\hline
  \end{tabularx}
\caption{Expansion coefficients for the invariants in the used model,
  Eq.~(\ref{bia}), up to third order in ${\bf k}$ which give rise to
  BIA spin splitting. All values in
eV\AA$^3$, except for $C_k$ in eV\AA.\cite{winklerbook}}
\label{BIA_coeff}
\end{table}
\bibliographystyle{apsrev4-1}
\bibliography{WK.bib}

\def\url#1{}
\begin{thebibliography}{45}%
\makeatletter
\providecommand \@ifxundefined [1]{%
 \@ifx{#1\undefined}
}%
\providecommand \@ifnum [1]{%
 \ifnum #1\expandafter \@firstoftwo
 \else \expandafter \@secondoftwo
 \fi
}%
\providecommand \@ifx [1]{%
 \ifx #1\expandafter \@firstoftwo
 \else \expandafter \@secondoftwo
 \fi
}%
\providecommand \natexlab [1]{#1}%
\providecommand \enquote  [1]{``#1''}%
\providecommand \bibnamefont  [1]{#1}%
\providecommand \bibfnamefont [1]{#1}%
\providecommand \citenamefont [1]{#1}%
\providecommand \href@noop [0]{\@secondoftwo}%
\providecommand \href [0]{\begingroup \@sanitize@url \@href}%
\providecommand \@href[1]{\@@startlink{#1}\@@href}%
\providecommand \@@href[1]{\endgroup#1\@@endlink}%
\providecommand \@sanitize@url [0]{\catcode `\\12\catcode `\$12\catcode
  `\&12\catcode `\#12\catcode `\^12\catcode `\_12\catcode `\%12\relax}%
\providecommand \@@startlink[1]{}%
\providecommand \@@endlink[0]{}%
\providecommand \url  [0]{\begingroup\@sanitize@url \@url }%
\providecommand \@url [1]{\endgroup\@href {#1}{\urlprefix }}%
\providecommand \urlprefix  [0]{URL }%
\providecommand \Eprint [0]{\href }%
\providecommand \doibase [0]{http://dx.doi.org/}%
\providecommand \selectlanguage [0]{\@gobble}%
\providecommand \bibinfo  [0]{\@secondoftwo}%
\providecommand \bibfield  [0]{\@secondoftwo}%
\providecommand \translation [1]{[#1]}%
\providecommand \BibitemOpen [0]{}%
\providecommand \bibitemStop [0]{}%
\providecommand \bibitemNoStop [0]{.\EOS\space}%
\providecommand \EOS [0]{\spacefactor3000\relax}%
\providecommand \BibitemShut  [1]{\csname bibitem#1\endcsname}%
\let\auto@bib@innerbib\@empty
\bibitem [{\citenamefont {Datta}\ and\ \citenamefont
  {Das}(1990)}]{:/content/aip/journal/apl/56/7/10.1063/1.102730}%
  \BibitemOpen
  \bibfield  {author} {\bibinfo {author} {\bibfnamefont {S.}~\bibnamefont
  {Datta}}\ and\ \bibinfo {author} {\bibfnamefont {B.}~\bibnamefont {Das}},\
  }\href {\doibase http://dx.doi.org/10.1063/1.102730} {\bibfield  {journal}
  {\bibinfo  {journal} {Appl. Phys. Lett.}\ }\textbf {\bibinfo {volume} {56}},\
  \bibinfo {pages} {665} (\bibinfo {year} {1990})}\BibitemShut {NoStop}%
\bibitem [{\citenamefont {D'yakonov}\ and\ \citenamefont
  {Perel'}(1971)}]{dyakonov72_3}%
  \BibitemOpen
  \bibfield  {author} {\bibinfo {author} {\bibfnamefont {M.~I.}\ \bibnamefont
  {D'yakonov}}\ and\ \bibinfo {author} {\bibfnamefont {V.~I.}\ \bibnamefont
  {Perel'}},\ }\href@noop {} {\bibfield  {journal} {\bibinfo  {journal} {Fiz.
  Tverd. Tela}\ }\textbf {\bibinfo {volume} {13}},\ \bibinfo {pages} {3581}
  (\bibinfo {year} {1971})},\ \bibinfo {note} {[Sov. Phys. Solid State {\bf
  13}, 3023 (1972)]}\BibitemShut {NoStop}%
\bibitem [{\citenamefont {Noether}(1918)}]{Noether1918}%
  \BibitemOpen
  \bibfield  {author} {\bibinfo {author} {\bibfnamefont {E.}~\bibnamefont
  {Noether}},\ }\href {http://eudml.org/doc/59024} {\bibfield  {journal}
  {\bibinfo  {journal} {Nachr. Konig. Gesellsch. Wiss. Gottingen, Math-Phys.
  Klasse}\ }\textbf {\bibinfo {volume} {1918}},\ \bibinfo {pages} {235}
  (\bibinfo {year} {1918})}\BibitemShut {NoStop}%
\bibitem [{\citenamefont {Dresselhaus}(1955)}]{Dresselhaus1955b}%
  \BibitemOpen
  \bibfield  {author} {\bibinfo {author} {\bibfnamefont {G.}~\bibnamefont
  {Dresselhaus}},\ }\href {\doibase 10.1103/PhysRev.100.580} {\bibfield
  {journal} {\bibinfo  {journal} {Phys. Rev.}\ }\textbf {\bibinfo {volume}
  {100}},\ \bibinfo {pages} {580} (\bibinfo {year} {1955})}\BibitemShut
  {NoStop}%
\bibitem [{\citenamefont {Rashba}(1960)}]{rashba_2}%
  \BibitemOpen
  \bibfield  {author} {\bibinfo {author} {\bibfnamefont {E.~I.}\ \bibnamefont
  {Rashba}},\ }\href@noop {} {\bibfield  {journal} {\bibinfo  {journal} {Sov.
  Phys.-Solid State}\ }\textbf {\bibinfo {volume} {2}},\ \bibinfo {pages}
  {1109} (\bibinfo {year} {1960})}\BibitemShut {NoStop}%
\bibitem [{\citenamefont {J.~Ohkawa}\ and\ \citenamefont
  {Uemura}(1974)}]{JPSJ.37.1325}%
  \BibitemOpen
  \bibfield  {author} {\bibinfo {author} {\bibfnamefont {F.}~\bibnamefont
  {J.~Ohkawa}}\ and\ \bibinfo {author} {\bibfnamefont {Y.}~\bibnamefont
  {Uemura}},\ }\href {\doibase 10.1143/JPSJ.37.1325} {\bibfield  {journal}
  {\bibinfo  {journal} {J. Phys. Soc. Jpn.}\ }\textbf {\bibinfo {volume}
  {37}},\ \bibinfo {pages} {1325} (\bibinfo {year} {1974})}\BibitemShut
  {NoStop}%
\bibitem [{\citenamefont {Schliemann}\ \emph {et~al.}(2003)\citenamefont
  {Schliemann}, \citenamefont {Egues},\ and\ \citenamefont
  {Loss}}]{Schliemann2003}%
  \BibitemOpen
  \bibfield  {author} {\bibinfo {author} {\bibfnamefont {J.}~\bibnamefont
  {Schliemann}}, \bibinfo {author} {\bibfnamefont {J.~C.}\ \bibnamefont
  {Egues}}, \ and\ \bibinfo {author} {\bibfnamefont {D.}~\bibnamefont {Loss}},\
  }\href {\doibase 10.1103/PhysRevLett.90.146801} {\bibfield  {journal}
  {\bibinfo  {journal} {Phys. Rev. Lett.}\ }\textbf {\bibinfo {volume} {90}},\
  \bibinfo {pages} {146801} (\bibinfo {year} {2003})}\BibitemShut {NoStop}%
\bibitem [{\citenamefont {Bernevig}\ \emph {et~al.}(2006)\citenamefont
  {Bernevig}, \citenamefont {Orenstein},\ and\ \citenamefont
  {Zhang}}]{Bernevig2006}%
  \BibitemOpen
  \bibfield  {author} {\bibinfo {author} {\bibfnamefont {B.~A.}\ \bibnamefont
  {Bernevig}}, \bibinfo {author} {\bibfnamefont {J.}~\bibnamefont {Orenstein}},
  \ and\ \bibinfo {author} {\bibfnamefont {S.-C.}\ \bibnamefont {Zhang}},\
  }\href {\doibase 10.1103/PhysRevLett.97.236601} {\bibfield  {journal}
  {\bibinfo  {journal} {Phys. Rev. Lett.}\ }\textbf {\bibinfo {volume} {97}},\
  \bibinfo {pages} {236601} (\bibinfo {year} {2006})}\BibitemShut {NoStop}%
\bibitem [{\citenamefont {Koralek}\ \emph {et~al.}(2009)\citenamefont
  {Koralek}, \citenamefont {Weber}, \citenamefont {Orenstein}, \citenamefont
  {Bernevig}, \citenamefont {Zhang}, \citenamefont {Mack},\ and\ \citenamefont
  {Awschalom}}]{Koralek2009}%
  \BibitemOpen
  \bibfield  {author} {\bibinfo {author} {\bibfnamefont {J.~D.}\ \bibnamefont
  {Koralek}}, \bibinfo {author} {\bibfnamefont {C.~P.}\ \bibnamefont {Weber}},
  \bibinfo {author} {\bibfnamefont {J.}~\bibnamefont {Orenstein}}, \bibinfo
  {author} {\bibfnamefont {B.~A.}\ \bibnamefont {Bernevig}}, \bibinfo {author}
  {\bibfnamefont {S.-C.}\ \bibnamefont {Zhang}}, \bibinfo {author}
  {\bibfnamefont {S.}~\bibnamefont {Mack}}, \ and\ \bibinfo {author}
  {\bibfnamefont {D.~D.}\ \bibnamefont {Awschalom}},\ }\href
  {http://dx.doi.org/10.1038/nature07871} {\bibfield  {journal} {\bibinfo
  {journal} {Nature}\ }\textbf {\bibinfo {volume} {458}},\ \bibinfo {pages}
  {610} (\bibinfo {year} {2009})}\BibitemShut {NoStop}%
\bibitem [{\citenamefont {Walser}\ \emph {et~al.}(2012)\citenamefont {Walser},
  \citenamefont {Reichl}, \citenamefont {Wegscheider},\ and\ \citenamefont
  {Salis}}]{10.1038/nphys2383}%
  \BibitemOpen
  \bibfield  {author} {\bibinfo {author} {\bibfnamefont {M.~P.}\ \bibnamefont
  {Walser}}, \bibinfo {author} {\bibfnamefont {C.}~\bibnamefont {Reichl}},
  \bibinfo {author} {\bibfnamefont {W.}~\bibnamefont {Wegscheider}}, \ and\
  \bibinfo {author} {\bibfnamefont {G.}~\bibnamefont {Salis}},\ }\href
  {\doibase http://dx.doi.org/10.1038/nphys2383} {\bibfield  {journal}
  {\bibinfo  {journal} {Nat Phys}\ }\textbf {\bibinfo {volume} {8}},\ \bibinfo
  {pages} {757} (\bibinfo {year} {2012})}\BibitemShut {NoStop}%
\bibitem [{\citenamefont {Rashba}\ and\ \citenamefont
  {Sherman}(1988)}]{Rashba1988175}%
  \BibitemOpen
  \bibfield  {author} {\bibinfo {author} {\bibfnamefont {E.}~\bibnamefont
  {Rashba}}\ and\ \bibinfo {author} {\bibfnamefont {E.}~\bibnamefont
  {Sherman}},\ }\href
  {http://www.sciencedirect.com/science/article/pii/0375960188901405}
  {\bibfield  {journal} {\bibinfo  {journal} {Phys. Lett. A}\ }\textbf
  {\bibinfo {volume} {129}},\ \bibinfo {pages} {175 } (\bibinfo {year}
  {1988})}\BibitemShut {NoStop}%
\bibitem [{\citenamefont {Sun}\ \emph {et~al.}(2010)\citenamefont {Sun},
  \citenamefont {Thompson},\ and\ \citenamefont
  {Nishida}}]{YSunThompsonNishidaBook}%
  \BibitemOpen
  \bibfield  {author} {\bibinfo {author} {\bibfnamefont {Y.}~\bibnamefont
  {Sun}}, \bibinfo {author} {\bibfnamefont {S.~E.}\ \bibnamefont {Thompson}}, \
  and\ \bibinfo {author} {\bibfnamefont {T.}~\bibnamefont {Nishida}},\
  }\href@noop {} {\emph {\bibinfo {title} {Strain Effects in Semiconductors:
  Theory and Device Applications}}}\ (\bibinfo  {publisher} {Springer},\
  \bibinfo {year} {2010})\BibitemShut {NoStop}%
\bibitem [{\citenamefont {Sun}\ \emph {et~al.}(2007)\citenamefont {Sun},
  \citenamefont {Thompson},\ and\ \citenamefont
  {Nishida}}]{:/content/aip/journal/jap/101/10/10.1063/1.2730561}%
  \BibitemOpen
  \bibfield  {author} {\bibinfo {author} {\bibfnamefont {Y.}~\bibnamefont
  {Sun}}, \bibinfo {author} {\bibfnamefont {S.~E.}\ \bibnamefont {Thompson}}, \
  and\ \bibinfo {author} {\bibfnamefont {T.}~\bibnamefont {Nishida}},\ }\href
  {\doibase http://dx.doi.org/10.1063/1.2730561} {\bibfield  {journal}
  {\bibinfo  {journal} {J. Appl. Phys.}\ }\textbf {\bibinfo {volume} {101}},\
  \bibinfo {eid} {104503} (\bibinfo {year} {2007})}\BibitemShut {NoStop}%
\bibitem [{\citenamefont {Seiler}\ \emph {et~al.}(1977)\citenamefont {Seiler},
  \citenamefont {Bajaj},\ and\ \citenamefont {Stephens}}]{Seiler77}%
  \BibitemOpen
  \bibfield  {author} {\bibinfo {author} {\bibfnamefont {D.~G.}\ \bibnamefont
  {Seiler}}, \bibinfo {author} {\bibfnamefont {B.~D.}\ \bibnamefont {Bajaj}}, \
  and\ \bibinfo {author} {\bibfnamefont {A.~E.}\ \bibnamefont {Stephens}},\
  }\href {\doibase 10.1103/PhysRevB.16.2822} {\bibfield  {journal} {\bibinfo
  {journal} {Phys. Rev. B}\ }\textbf {\bibinfo {volume} {16}},\ \bibinfo
  {pages} {2822} (\bibinfo {year} {1977})}\BibitemShut {NoStop}%
\bibitem [{\citenamefont {Cardona}\ \emph {et~al.}(1988)\citenamefont
  {Cardona}, \citenamefont {Christensen},\ and\ \citenamefont
  {Fasol}}]{PhysRevB.38.1806}%
  \BibitemOpen
  \bibfield  {author} {\bibinfo {author} {\bibfnamefont {M.}~\bibnamefont
  {Cardona}}, \bibinfo {author} {\bibfnamefont {N.~E.}\ \bibnamefont
  {Christensen}}, \ and\ \bibinfo {author} {\bibfnamefont {G.}~\bibnamefont
  {Fasol}},\ }\href {\doibase 10.1103/PhysRevB.38.1806} {\bibfield  {journal}
  {\bibinfo  {journal} {Phys. Rev. B}\ }\textbf {\bibinfo {volume} {38}},\
  \bibinfo {pages} {1806} (\bibinfo {year} {1988})}\BibitemShut {NoStop}%
\bibitem [{\citenamefont {Korn}\ \emph {et~al.}(2010)\citenamefont {Korn},
  \citenamefont {Kugler}, \citenamefont {Griesbeck}, \citenamefont {Schulz},
  \citenamefont {Wagner}, \citenamefont {Hirmer}, \citenamefont {Gerl},
  \citenamefont {Schuh}, \citenamefont {Wegscheider},\ and\ \citenamefont
  {Sch\"uller}}]{1367-2630-12-4-043003}%
  \BibitemOpen
  \bibfield  {author} {\bibinfo {author} {\bibfnamefont {T.}~\bibnamefont
  {Korn}}, \bibinfo {author} {\bibfnamefont {M.}~\bibnamefont {Kugler}},
  \bibinfo {author} {\bibfnamefont {M.}~\bibnamefont {Griesbeck}}, \bibinfo
  {author} {\bibfnamefont {R.}~\bibnamefont {Schulz}}, \bibinfo {author}
  {\bibfnamefont {A.}~\bibnamefont {Wagner}}, \bibinfo {author} {\bibfnamefont
  {M.}~\bibnamefont {Hirmer}}, \bibinfo {author} {\bibfnamefont
  {C.}~\bibnamefont {Gerl}}, \bibinfo {author} {\bibfnamefont {D.}~\bibnamefont
  {Schuh}}, \bibinfo {author} {\bibfnamefont {W.}~\bibnamefont {Wegscheider}},
  \ and\ \bibinfo {author} {\bibfnamefont {C.}~\bibnamefont {Sch\"uller}},\
  }\href {http://stacks.iop.org/1367-2630/12/i=4/a=043003} {\bibfield
  {journal} {\bibinfo  {journal} {New J. Phys.}\ }\textbf {\bibinfo {volume}
  {12}},\ \bibinfo {pages} {043003} (\bibinfo {year} {2010})}\BibitemShut
  {NoStop}%
\bibitem [{\citenamefont {Kugler}\ \emph {et~al.}(2009)\citenamefont {Kugler},
  \citenamefont {Andlauer}, \citenamefont {Korn}, \citenamefont {Wagner},
  \citenamefont {Fehringer}, \citenamefont {Schulz}, \citenamefont {Kubov\'a},
  \citenamefont {Gerl}, \citenamefont {Schuh}, \citenamefont {Wegscheider},
  \citenamefont {Vogl},\ and\ \citenamefont {Sch\"uller}}]{PhysRevB.80.035325}%
  \BibitemOpen
  \bibfield  {author} {\bibinfo {author} {\bibfnamefont {M.}~\bibnamefont
  {Kugler}}, \bibinfo {author} {\bibfnamefont {T.}~\bibnamefont {Andlauer}},
  \bibinfo {author} {\bibfnamefont {T.}~\bibnamefont {Korn}}, \bibinfo {author}
  {\bibfnamefont {A.}~\bibnamefont {Wagner}}, \bibinfo {author} {\bibfnamefont
  {S.}~\bibnamefont {Fehringer}}, \bibinfo {author} {\bibfnamefont
  {R.}~\bibnamefont {Schulz}}, \bibinfo {author} {\bibfnamefont
  {M.}~\bibnamefont {Kubov\'a}}, \bibinfo {author} {\bibfnamefont
  {C.}~\bibnamefont {Gerl}}, \bibinfo {author} {\bibfnamefont {D.}~\bibnamefont
  {Schuh}}, \bibinfo {author} {\bibfnamefont {W.}~\bibnamefont {Wegscheider}},
  \bibinfo {author} {\bibfnamefont {P.}~\bibnamefont {Vogl}}, \ and\ \bibinfo
  {author} {\bibfnamefont {C.}~\bibnamefont {Sch\"uller}},\ }\href {\doibase
  10.1103/PhysRevB.80.035325} {\bibfield  {journal} {\bibinfo  {journal} {Phys.
  Rev. B}\ }\textbf {\bibinfo {volume} {80}},\ \bibinfo {pages} {035325}
  (\bibinfo {year} {2009})}\BibitemShut {NoStop}%
\bibitem [{\citenamefont {Hirmer}\ \emph {et~al.}(2011)\citenamefont {Hirmer},
  \citenamefont {Hirmer}, \citenamefont {Schuh}, \citenamefont {Wegscheider},
  \citenamefont {Korn}, \citenamefont {Winkler},\ and\ \citenamefont
  {Sch\"uller}}]{PhysRevLett.107.216805}%
  \BibitemOpen
  \bibfield  {author} {\bibinfo {author} {\bibfnamefont {M.}~\bibnamefont
  {Hirmer}}, \bibinfo {author} {\bibfnamefont {M.}~\bibnamefont {Hirmer}},
  \bibinfo {author} {\bibfnamefont {D.}~\bibnamefont {Schuh}}, \bibinfo
  {author} {\bibfnamefont {W.}~\bibnamefont {Wegscheider}}, \bibinfo {author}
  {\bibfnamefont {T.}~\bibnamefont {Korn}}, \bibinfo {author} {\bibfnamefont
  {R.}~\bibnamefont {Winkler}}, \ and\ \bibinfo {author} {\bibfnamefont
  {C.}~\bibnamefont {Sch\"uller}},\ }\href {\doibase
  10.1103/PhysRevLett.107.216805} {\bibfield  {journal} {\bibinfo  {journal}
  {Phys. Rev. Lett.}\ }\textbf {\bibinfo {volume} {107}},\ \bibinfo {pages}
  {216805} (\bibinfo {year} {2011})}\BibitemShut {NoStop}%
\bibitem [{\citenamefont {Grbi\'{c}}\ \emph {et~al.}(2008)\citenamefont
  {Grbi\'{c}}, \citenamefont {Leturcq}, \citenamefont {Ihn}, \citenamefont
  {Ensslin}, \citenamefont {Reuter},\ and\ \citenamefont
  {Wieck}}]{PhysRevB.77.125312}%
  \BibitemOpen
  \bibfield  {author} {\bibinfo {author} {\bibfnamefont {B.}~\bibnamefont
  {Grbi\'{c}}}, \bibinfo {author} {\bibfnamefont {R.}~\bibnamefont {Leturcq}},
  \bibinfo {author} {\bibfnamefont {T.}~\bibnamefont {Ihn}}, \bibinfo {author}
  {\bibfnamefont {K.}~\bibnamefont {Ensslin}}, \bibinfo {author} {\bibfnamefont
  {D.}~\bibnamefont {Reuter}}, \ and\ \bibinfo {author} {\bibfnamefont {A.~D.}\
  \bibnamefont {Wieck}},\ }\href {\doibase 10.1103/PhysRevB.77.125312}
  {\bibfield  {journal} {\bibinfo  {journal} {Phys. Rev. B}\ }\textbf {\bibinfo
  {volume} {77}},\ \bibinfo {pages} {125312} (\bibinfo {year}
  {2008})}\BibitemShut {NoStop}%
\bibitem [{\citenamefont {Sacksteder}\ and\ \citenamefont
  {Bernevig}(2014)}]{PhysRevB.89.161307}%
  \BibitemOpen
  \bibfield  {author} {\bibinfo {author} {\bibfnamefont {V.~E.}\ \bibnamefont
  {Sacksteder}}\ and\ \bibinfo {author} {\bibfnamefont {B.~A.}\ \bibnamefont
  {Bernevig}},\ }\href {\doibase 10.1103/PhysRevB.89.161307} {\bibfield
  {journal} {\bibinfo  {journal} {Phys. Rev. B}\ }\textbf {\bibinfo {volume}
  {89}},\ \bibinfo {pages} {161307} (\bibinfo {year} {2014})}\BibitemShut
  {NoStop}%
\bibitem [{\citenamefont {Dollinger}\ \emph {et~al.}(2014)\citenamefont
  {Dollinger}, \citenamefont {Kammermeier}, \citenamefont {Scholz},
  \citenamefont {Wenk}, \citenamefont {Schliemann}, \citenamefont {Richter},\
  and\ \citenamefont {Winkler}}]{PhysRevB.90.115306}%
  \BibitemOpen
  \bibfield  {author} {\bibinfo {author} {\bibfnamefont {T.}~\bibnamefont
  {Dollinger}}, \bibinfo {author} {\bibfnamefont {M.}~\bibnamefont
  {Kammermeier}}, \bibinfo {author} {\bibfnamefont {A.}~\bibnamefont {Scholz}},
  \bibinfo {author} {\bibfnamefont {P.}~\bibnamefont {Wenk}}, \bibinfo {author}
  {\bibfnamefont {J.}~\bibnamefont {Schliemann}}, \bibinfo {author}
  {\bibfnamefont {K.}~\bibnamefont {Richter}}, \ and\ \bibinfo {author}
  {\bibfnamefont {R.}~\bibnamefont {Winkler}},\ }\href {\doibase
  10.1103/PhysRevB.90.115306} {\bibfield  {journal} {\bibinfo  {journal} {Phys.
  Rev. B}\ }\textbf {\bibinfo {volume} {90}},\ \bibinfo {pages} {115306}
  (\bibinfo {year} {2014})}\BibitemShut {NoStop}%
\bibitem [{\citenamefont {Vurgaftman}\ \emph {et~al.}(2001)\citenamefont
  {Vurgaftman}, \citenamefont {Meyer},\ and\ \citenamefont
  {Ram-Mohan}}]{:/content/aip/journal/jap/89/11/10.1063/1.1368156}%
  \BibitemOpen
  \bibfield  {author} {\bibinfo {author} {\bibfnamefont {I.}~\bibnamefont
  {Vurgaftman}}, \bibinfo {author} {\bibfnamefont {J.~R.}\ \bibnamefont
  {Meyer}}, \ and\ \bibinfo {author} {\bibfnamefont {L.~R.}\ \bibnamefont
  {Ram-Mohan}},\ }\href {\doibase http://dx.doi.org/10.1063/1.1368156}
  {\bibfield  {journal} {\bibinfo  {journal} {J. Appl. Phys.}\ }\textbf
  {\bibinfo {volume} {89}},\ \bibinfo {pages} {5815} (\bibinfo {year}
  {2001})}\BibitemShut {NoStop}%
\bibitem [{\citenamefont {Yu}\ and\ \citenamefont
  {Cardona}(2010)}]{YuCardonaBook}%
  \BibitemOpen
  \bibfield  {author} {\bibinfo {author} {\bibfnamefont {P.}~\bibnamefont
  {Yu}}\ and\ \bibinfo {author} {\bibfnamefont {M.}~\bibnamefont {Cardona}},\
  }\href@noop {} {\emph {\bibinfo {title} {Fundamentals of Semiconductors:
  Physics and Materials Properties}}},\ Graduate Texts in Physics\ (\bibinfo
  {publisher} {Springer},\ \bibinfo {year} {2010})\BibitemShut {NoStop}%
\bibitem [{\citenamefont {Winkler}(2003)}]{winklerbook}%
  \BibitemOpen
  \bibfield  {author} {\bibinfo {author} {\bibfnamefont {R.}~\bibnamefont
  {Winkler}},\ }\href@noop {} {\emph {\bibinfo {title} {{Spin-Orbit Coupling
  Effects in Two-Dimensional Electron and Hole Systems}}}},\ \bibinfo {series}
  {Springer Tracts in Modern Physics}, Vol.\ \bibinfo {volume} {191}\ (\bibinfo
   {publisher} {Springer-Verlag, Berlin},\ \bibinfo {year} {2003})\BibitemShut
  {NoStop}%
\bibitem [{Note1()}]{Note1}%
  \BibitemOpen
  \bibinfo {note} {A negative value of $\gamma _2/\gamma _3$ appears, e.g.,
  when the lowest conduction band is not an s-type band as it is the case in
  diamond\cite {PhysRevB.50.18054}. However, for diamond one finds only $\gamma
  _2/\gamma _3\approx -0.16$\cite {PhysRevB.50.18054}.}\BibitemShut {Stop}%
\bibitem [{\citenamefont {Bulaev}\ and\ \citenamefont
  {Loss}(2005)}]{PhysRevLett.95.076805}%
  \BibitemOpen
  \bibfield  {author} {\bibinfo {author} {\bibfnamefont {D.~V.}\ \bibnamefont
  {Bulaev}}\ and\ \bibinfo {author} {\bibfnamefont {D.}~\bibnamefont {Loss}},\
  }\href {\doibase 10.1103/PhysRevLett.95.076805} {\bibfield  {journal}
  {\bibinfo  {journal} {Phys. Rev. Lett.}\ }\textbf {\bibinfo {volume} {95}},\
  \bibinfo {pages} {076805} (\bibinfo {year} {2005})}\BibitemShut {NoStop}%
\bibitem [{\citenamefont {Luo}\ \emph {et~al.}(2010)\citenamefont {Luo},
  \citenamefont {Chantis}, \citenamefont {van Schilfgaarde}, \citenamefont
  {Bester},\ and\ \citenamefont {Zunger}}]{PhysRevLett.104.066405}%
  \BibitemOpen
  \bibfield  {author} {\bibinfo {author} {\bibfnamefont {J.-W.}\ \bibnamefont
  {Luo}}, \bibinfo {author} {\bibfnamefont {A.~N.}\ \bibnamefont {Chantis}},
  \bibinfo {author} {\bibfnamefont {M.}~\bibnamefont {van Schilfgaarde}},
  \bibinfo {author} {\bibfnamefont {G.}~\bibnamefont {Bester}}, \ and\ \bibinfo
  {author} {\bibfnamefont {A.}~\bibnamefont {Zunger}},\ }\href {\doibase
  10.1103/PhysRevLett.104.066405} {\bibfield  {journal} {\bibinfo  {journal}
  {Phys. Rev. Lett.}\ }\textbf {\bibinfo {volume} {104}},\ \bibinfo {pages}
  {066405} (\bibinfo {year} {2010})}\BibitemShut {NoStop}%
\bibitem [{\citenamefont {Durnev}\ \emph {et~al.}(2014)\citenamefont {Durnev},
  \citenamefont {Glazov},\ and\ \citenamefont {Ivchenko}}]{PhysRevB.89.075430}%
  \BibitemOpen
  \bibfield  {author} {\bibinfo {author} {\bibfnamefont {M.~V.}\ \bibnamefont
  {Durnev}}, \bibinfo {author} {\bibfnamefont {M.~M.}\ \bibnamefont {Glazov}},
  \ and\ \bibinfo {author} {\bibfnamefont {E.~L.}\ \bibnamefont {Ivchenko}},\
  }\href {\doibase 10.1103/PhysRevB.89.075430} {\bibfield  {journal} {\bibinfo
  {journal} {Phys. Rev. B}\ }\textbf {\bibinfo {volume} {89}},\ \bibinfo
  {pages} {075430} (\bibinfo {year} {2014})}\BibitemShut {NoStop}%
\bibitem [{\citenamefont {Luttinger}(1956)}]{Luttinger1956}%
  \BibitemOpen
  \bibfield  {author} {\bibinfo {author} {\bibfnamefont {J.~M.}\ \bibnamefont
  {Luttinger}},\ }\href {http://dx.doi.org/10.1103/PhysRev.102.1030} {\bibfield
   {journal} {\bibinfo  {journal} {Phys. Rev.}\ }\textbf {\bibinfo {volume}
  {102}},\ \bibinfo {pages} {1030} (\bibinfo {year} {1956})}\BibitemShut
  {NoStop}%
\bibitem [{\citenamefont {Bir}\ and\ \citenamefont
  {Pikus}(1974)}]{BirPikusSymmetryStrain1974}%
  \BibitemOpen
  \bibfield  {author} {\bibinfo {author} {\bibfnamefont {G.~L.}\ \bibnamefont
  {Bir}}\ and\ \bibinfo {author} {\bibfnamefont {G.~E.}\ \bibnamefont
  {Pikus}},\ }\href@noop {} {\emph {\bibinfo {title} {{Symmetry and
  Strain-Induced Effects in Semiconductors}}}}\ (\bibinfo  {publisher}
  {Wiley/Halsted Press},\ \bibinfo {year} {1974})\BibitemShut {NoStop}%
\bibitem [{\citenamefont {Ivchenko}\ and\ \citenamefont
  {Pikus}(1995)}]{IvchenkoSuperlattices}%
  \BibitemOpen
  \bibfield  {author} {\bibinfo {author} {\bibfnamefont {E.}~\bibnamefont
  {Ivchenko}}\ and\ \bibinfo {author} {\bibfnamefont {P.}~\bibnamefont
  {Pikus}},\ }\href@noop {} {\emph {\bibinfo {title} {Superlattices and Other
  Heterostructures: Symmetry and Optical Phenomena}}},\ Springer Series in
  Solid-State Science 110\ (\bibinfo  {publisher} {Springer},\ \bibinfo {year}
  {1995})\BibitemShut {NoStop}%
\bibitem [{\citenamefont {Ivchenko}\ \emph {et~al.}(1996)\citenamefont
  {Ivchenko}, \citenamefont {Kaminski},\ and\ \citenamefont
  {R\"ossler}}]{PhysRevB.54.5852}%
  \BibitemOpen
  \bibfield  {author} {\bibinfo {author} {\bibfnamefont {E.~L.}\ \bibnamefont
  {Ivchenko}}, \bibinfo {author} {\bibfnamefont {A.~Y.}\ \bibnamefont
  {Kaminski}}, \ and\ \bibinfo {author} {\bibfnamefont {U.}~\bibnamefont
  {R\"ossler}},\ }\href {\doibase 10.1103/PhysRevB.54.5852} {\bibfield
  {journal} {\bibinfo  {journal} {Phys. Rev. B}\ }\textbf {\bibinfo {volume}
  {54}},\ \bibinfo {pages} {5852} (\bibinfo {year} {1996})}\BibitemShut
  {NoStop}%
\bibitem [{\citenamefont {Magri}\ and\ \citenamefont
  {Zunger}(2000)}]{PhysRevB.62.10364}%
  \BibitemOpen
  \bibfield  {author} {\bibinfo {author} {\bibfnamefont {R.}~\bibnamefont
  {Magri}}\ and\ \bibinfo {author} {\bibfnamefont {A.}~\bibnamefont {Zunger}},\
  }\href {\doibase 10.1103/PhysRevB.62.10364} {\bibfield  {journal} {\bibinfo
  {journal} {Phys. Rev. B}\ }\textbf {\bibinfo {volume} {62}},\ \bibinfo
  {pages} {10364} (\bibinfo {year} {2000})}\BibitemShut {NoStop}%
\bibitem [{Note2()}]{Note2}%
  \BibitemOpen
  \bibinfo {note} {Note that the definition in Eq.~(\ref {beta_def}) differs
  from the one given in Ref.~\protect \rev@citealpnum {PhysRevB.89.161307} in
  the term proportional to $d$ by a factor of $1/\protect \sqrt
  {3}$.}\BibitemShut {Stop}%
\bibitem [{\citenamefont {Kane}(1957)}]{Kane1957249}%
  \BibitemOpen
  \bibfield  {author} {\bibinfo {author} {\bibfnamefont {E.~O.}\ \bibnamefont
  {Kane}},\ }\href {\doibase http://dx.doi.org/10.1016/0022-3697(57)90013-6}
  {\bibfield  {journal} {\bibinfo  {journal} {J. Phys. Chem. Solids}\ }\textbf
  {\bibinfo {volume} {1}},\ \bibinfo {pages} {249 } (\bibinfo {year}
  {1957})}\BibitemShut {NoStop}%
\bibitem [{\citenamefont {Habib}\ \emph {et~al.}(2007)\citenamefont {Habib},
  \citenamefont {Shabani}, \citenamefont {De~Poortere}, \citenamefont
  {Shayegan},\ and\ \citenamefont {Winkler}}]{Habib2007}%
  \BibitemOpen
  \bibfield  {author} {\bibinfo {author} {\bibfnamefont {B.}~\bibnamefont
  {Habib}}, \bibinfo {author} {\bibfnamefont {J.}~\bibnamefont {Shabani}},
  \bibinfo {author} {\bibfnamefont {E.~P.}\ \bibnamefont {De~Poortere}},
  \bibinfo {author} {\bibfnamefont {M.}~\bibnamefont {Shayegan}}, \ and\
  \bibinfo {author} {\bibfnamefont {R.}~\bibnamefont {Winkler}},\ }\href
  {\doibase 10.1103/PhysRevB.75.153304} {\bibfield  {journal} {\bibinfo
  {journal} {Phys. Rev. B}\ }\textbf {\bibinfo {volume} {75}},\ \bibinfo
  {pages} {153304} (\bibinfo {year} {2007})}\BibitemShut {NoStop}%
\bibitem [{\citenamefont {Knap}\ \emph {et~al.}(1996)\citenamefont {Knap},
  \citenamefont {Skierbiszewski}, \citenamefont {Zduniak}, \citenamefont
  {Litwin-Staszewska}, \citenamefont {Bertho}, \citenamefont {Kobbi},
  \citenamefont {Robert}, \citenamefont {Pikus}, \citenamefont {Pikus},
  \citenamefont {Iordanskii}, \citenamefont {Mosser}, \citenamefont
  {Zekentes},\ and\ \citenamefont {Lyanda-Geller}}]{Knap1996}%
  \BibitemOpen
  \bibfield  {author} {\bibinfo {author} {\bibfnamefont {W.}~\bibnamefont
  {Knap}}, \bibinfo {author} {\bibfnamefont {C.}~\bibnamefont
  {Skierbiszewski}}, \bibinfo {author} {\bibfnamefont {A.}~\bibnamefont
  {Zduniak}}, \bibinfo {author} {\bibfnamefont {E.}~\bibnamefont
  {Litwin-Staszewska}}, \bibinfo {author} {\bibfnamefont {D.}~\bibnamefont
  {Bertho}}, \bibinfo {author} {\bibfnamefont {F.}~\bibnamefont {Kobbi}},
  \bibinfo {author} {\bibfnamefont {J.~L.}\ \bibnamefont {Robert}}, \bibinfo
  {author} {\bibfnamefont {G.~E.}\ \bibnamefont {Pikus}}, \bibinfo {author}
  {\bibfnamefont {F.~G.}\ \bibnamefont {Pikus}}, \bibinfo {author}
  {\bibfnamefont {S.~V.}\ \bibnamefont {Iordanskii}}, \bibinfo {author}
  {\bibfnamefont {V.}~\bibnamefont {Mosser}}, \bibinfo {author} {\bibfnamefont
  {K.}~\bibnamefont {Zekentes}}, \ and\ \bibinfo {author} {\bibfnamefont
  {Y.~B.}\ \bibnamefont {Lyanda-Geller}},\ }\href
  {http://link.aps.org/doi/10.1103/PhysRevB.53.3912} {\bibfield  {journal}
  {\bibinfo  {journal} {Phys. Rev. B}\ }\textbf {\bibinfo {volume} {53}},\
  \bibinfo {pages} {3912} (\bibinfo {year} {1996})}\BibitemShut {NoStop}%
\bibitem [{\citenamefont {Dollinger}(2013)}]{DollingerThesis2013}%
  \BibitemOpen
  \bibfield  {author} {\bibinfo {author} {\bibfnamefont {T.}~\bibnamefont
  {Dollinger}},\ }\emph {\bibinfo {title} {Spin Transport in Two-Dimensional
  Electron and Hole Gases}},\ \href
  {http://epub.uni-regensburg.de/29909/1/dollinger.pdf} {Ph.D. thesis},\
  \bibinfo  {school} {Universit\"at Regensburg} (\bibinfo {year}
  {2013})\BibitemShut {NoStop}%
\bibitem [{\citenamefont {Kettemann}(2007)}]{Kettemann:PRL98:2007}%
  \BibitemOpen
  \bibfield  {author} {\bibinfo {author} {\bibfnamefont {S.}~\bibnamefont
  {Kettemann}},\ }\href {\doibase 10.1103/PhysRevLett.98.176808} {\bibfield
  {journal} {\bibinfo  {journal} {Phys. Rev. Lett.}\ }\textbf {\bibinfo
  {volume} {98}},\ \bibinfo {pages} {176808} (\bibinfo {year}
  {2007})}\BibitemShut {NoStop}%
\bibitem [{Note3()}]{Note3}%
  \BibitemOpen
  \bibinfo {note} {Eq.~(\ref {gamma3gamma2}) is valid if the reduced Luttinger
  parameters $\gamma _2^\prime $ and $\gamma _3^\prime $ in the applied model
  vanish\cite {Mayer1991}.}\BibitemShut {Stop}%
\bibitem [{Note4()}]{Note4}%
  \BibitemOpen
  \bibinfo {note} {Following the Koster notation\cite {Koster1957173}, one has
  for the tetrahedral point group $T_d$ three double group representations
  given by the two two-dimensional representations $\Gamma _6$, $\Gamma _7$ and
  the four dimensional one $\Gamma _8$.}\BibitemShut {Stop}%
\bibitem [{\citenamefont {Pikus}\ \emph {et~al.}(1988)\citenamefont {Pikus},
  \citenamefont {Maruschak},\ and\ \citenamefont
  {Titkov}}]{PikusTitkov23.1988}%
  \BibitemOpen
  \bibfield  {author} {\bibinfo {author} {\bibfnamefont {G.}~\bibnamefont
  {Pikus}}, \bibinfo {author} {\bibfnamefont {V.}~\bibnamefont {Maruschak}}, \
  and\ \bibinfo {author} {\bibfnamefont {A.}~\bibnamefont {Titkov}},\
  }\href@noop {} {\bibfield  {journal} {\bibinfo  {journal} {Fiz. Tekh.
  Poluprovodn.}\ }\textbf {\bibinfo {volume} {23}} (\bibinfo {year}
  {1988})}\BibitemShut {NoStop}%
\bibitem [{\citenamefont {Willatzen}\ \emph {et~al.}(1994)\citenamefont
  {Willatzen}, \citenamefont {Cardona},\ and\ \citenamefont
  {Christensen}}]{PhysRevB.50.18054}%
  \BibitemOpen
  \bibfield  {author} {\bibinfo {author} {\bibfnamefont {M.}~\bibnamefont
  {Willatzen}}, \bibinfo {author} {\bibfnamefont {M.}~\bibnamefont {Cardona}},
  \ and\ \bibinfo {author} {\bibfnamefont {N.~E.}\ \bibnamefont
  {Christensen}},\ }\href {\doibase 10.1103/PhysRevB.50.18054} {\bibfield
  {journal} {\bibinfo  {journal} {Phys. Rev. B}\ }\textbf {\bibinfo {volume}
  {50}},\ \bibinfo {pages} {18054} (\bibinfo {year} {1994})}\BibitemShut
  {NoStop}%
\bibitem [{\citenamefont {Mayer}\ and\ \citenamefont
  {R\"{o}ssler}(1991)}]{Mayer1991}%
  \BibitemOpen
  \bibfield  {author} {\bibinfo {author} {\bibfnamefont {H.}~\bibnamefont
  {Mayer}}\ and\ \bibinfo {author} {\bibfnamefont {U.}~\bibnamefont
  {R\"{o}ssler}},\ }\href {\doibase 10.1103/PhysRevB.44.9048} {\bibfield
  {journal} {\bibinfo  {journal} {Physical Review B}\ }\textbf {\bibinfo
  {volume} {44}},\ \bibinfo {pages} {9048} (\bibinfo {year}
  {1991})}\BibitemShut {NoStop}%
\bibitem [{\citenamefont {Koster}(1957)}]{Koster1957173}%
  \BibitemOpen
  \bibfield  {author} {\bibinfo {author} {\bibfnamefont {G.}~\bibnamefont
  {Koster}},\ }in\ \href {\doibase
  http://dx.doi.org/10.1016/S0081-1947(08)60103-4} {\emph {\bibinfo {booktitle}
  {Solid State Physics}}},\ Vol.~\bibinfo {volume} {5},\ \bibinfo {editor}
  {edited by\ \bibinfo {editor} {\bibfnamefont {F.}~\bibnamefont {Seitz}}\ and\
  \bibinfo {editor} {\bibfnamefont {D.}~\bibnamefont {Turnbull}}}\ (\bibinfo
  {publisher} {Academic Press},\ \bibinfo {year} {1957})\ pp.\ \bibinfo {pages}
  {173 -- 256}\BibitemShut {NoStop}%
\end{thebibliography}%
\end{document}